\documentclass[superscriptaddress, amsmath, amssymb, aps, prb, floatfix, twocolumn]{revtex4-2}
\pdfoutput=1
\usepackage[utf8]{inputenc}
\usepackage[T1]{fontenc}
\usepackage{babel}
\usepackage{graphicx}
\usepackage{amsfonts}
\usepackage{amssymb}
\usepackage{amsmath}
\usepackage{amsthm}
\usepackage{mathtools}
\usepackage{physics}
\usepackage[normalem]{ulem}
\usepackage{blindtext}
\usepackage{dsfont}

\def\id{{\rm 1\kern-.22em l}}

\usepackage{mathtools}
\usepackage{braket}
\renewcommand\bra[1]{{\langle{#1}|}}
\renewcommand\ket[1]{{|{#1}\rangle}}
\usepackage{graphicx}
\usepackage[format=plain,justification=centerlast]{caption}
\usepackage[format=plain,justification=centerlast]{subcaption}
\usepackage{bm}
\usepackage[dvipsnames]{xcolor}

\usepackage{natbib}
\setcitestyle{square,numbers}
\usepackage[colorlinks]{hyperref}

\begin{document}

\title{Generalization of Gisin's Theorem to Quantum Fields} 
\author{Konrad Schlichtholz}
\email[]{konrad.schlichtholz@phdstud.ug.edu.pl}
\affiliation{International Centre for Theory of Quantum Technologies, University of Gdansk, 80-309 Gda{\'n}sk, Poland}
\author{Marcin Markiewicz}
\email[]{marcin.markiewicz@ug.edu.pl}
\affiliation{International Centre for Theory of Quantum Technologies, University of Gdansk, 80-309 Gda{\'n}sk, Poland}
\pagenumbering{arabic}

\begin{abstract}
We generalize Gisin's theorem on the relation between the entanglement of pure states and Bell non-classicality to the case of mode entanglement of separated groups of modes of quantum fields, extending the theorem to cover also states with undefined particle number. We show that \textit{any} pure state of the field which contains entanglement between two groups of separated modes violates some Clauser-Horne inequality.  In order to construct the observables leading to a violation in the first step, we show an isomorphism between the Fock space built from a single-particle space involving two separated groups of modes and a tensor product of two abstract separable Hilbert spaces spanned by formal monomials of creation operators. In the second step, we perform a Schmidt decomposition of a given entangled state mapped to this tensor product space and then we map back the obtained Schmidt decomposition to the original Fock space of the system under consideration. Such obtained Schmidt decomposition in Fock space allows for the construction of observables leading to a violation of the Clauser-Horne inequality. We also show that our generalization of Gisin's theorem holds for the case of states on non-separable Hilbert spaces, which physically represent states with actually infinite number of particles. Such states emerge, for example, in the discussion of quantum phase transitions. Finally, we discuss the experimental feasibility of the constructed Bell test and provide a necessary condition for the realizability of this test within the realm of passive linear optics.
\end{abstract}

\maketitle
\section{introduction}
The superposition principle is one of the quantum phenomena that has impacted our fundamental understanding of nature the most. Especially the  concept of entanglement, which emerges from the superposition principle, exposes the failure of classical intuitions like local realism presented in the EPR paradox \cite{EPR} and discussed in the entire Bell-inequalities literature \cite{Aspect2002,Brukner2012,Pan2012,Werner2001,Zukowski2002,BrunnerRev}.  The introduction of Bell inequalities \cite{BELL} allowed for direct proof of the incompatibility of quantum mechanics with local hidden variable  (LHV) probabilistic models giving rise to the phenomenon of Bell non-classicality.  Those theoretical predictions were then experimentally tested confirming the predictions of quantum mechanics \cite{Aspect1982,Hensen2015}. 

This confirmation, however, does not answer the question whether Bell non-classicality is a general feature of pure entangled states. Especially the question whether Bell non-classicality can also be revealed by arbitrarily weakly entangled states. The first step in answering this question was formulation of Gisin's theorem \cite{Gisin_PLA} which states that any entangled \textit{two-qubit} state violates some Bell inequality. This was followed by an extension to a higher number of systems \cite{POPESCU} with recent complement \cite{GACHECHILADZE}.  Theorem was further generalized to an arbitrarily large number of  particles and the arbitrary dimension of single particle Hilbert spaces \cite{Gisin_inf2,Gisin_inf}. However, those generalizations do not provide a full answer to the question, since the analysis therein is performed in terms of first quantization, which requires fixing the number of particles. However, it is known that entanglement also exists for states with an undefined number of particles \cite{Masza, two_mode_ent, Hardy94,Enk05}, which requires the second quantization for its description. Therefore, the natural framework for treating this problem is provided by quantum field theory which more closely describes nature and allows for better insight into foundations of physical theories.

In this paper, we present a generalization of Gisin's theorem to the arbitrary entangled pure state in quantum field theory. We present the construction of the Schmidt decomposition for multimode states in the Fock space, which then leads to the construction of the state-specific Clauser-Horne (CH) inequality \cite{CH} violated by the given arbitrary entangled state.
We show that our generalization holds also in the case of a non-separable extension of Fock space (which includes states with infinite number of particles) in terms of entanglement of quasi-particles. Finally, we show the necessary condition for Bell non-classicality to be observed by projective particle-number-non-mixing measurements realizable within passive linear optics. This condition allows for determining whether the homodyne measurement or another non-projective POVM-type measurement is necessary to demonstrate Bell non-classicality for a given mode-entangled state.

\section{Preliminaries}

In order to consider entanglement, one has to introduce the state space of the theory.
Let us briefly recall the construction of the state space in the first and second quantization. 

\subsection{First quantization state space and entanglement}
The state of a single particle in the first quantization is specified by a normalized vector $\ket{\psi}$ in the single particle Hilbert space $\mathcal{H}$. This Hilbert space, in general, is isomorphic to the space of square-integrable functions  $L^2(\mathcal{M})$ on a given $d$-dimensional manifold  $\mathcal{M}$ in which the particle lives or on its subspace determined by the evolution equations.  Therefore, $\mathcal{H}$ is a separable Hilbert space, which is equivalent to saying that one can construct a countable basis in $\mathcal{H}$. Then, the state space of two distinct particles is given by the tensor product of the corresponding single particle spaces $\mathcal{H}_1\otimes\mathcal{H}_2$. Such construction gives rise to a natural division of the system into two subsystems among which the entanglement can be considered. In general, in the case of two separable Hilbert spaces constructed by means of a tensor product, one can decompose any state $\ket{\psi_{12}}\in \mathcal{H}_1\otimes\mathcal{H}_2$ in terms of Schmidt decomposition \cite{reed1981}:
\begin{equation}
    \ket{\psi_{12}}=\sum_t \lambda_t \ket{\phi_t}_1\otimes \ket{\varphi_t}_2,
\end{equation}
where $\sum_t |\lambda_t|^2=1$, 
 $\braket{\phi_t|\phi_{t'}}=\braket{\varphi_t|\varphi_{t'}}=\delta_{t,t'}$ and  $\delta_{t,t'}$ stands for the Kronecker delta. Upon this decomposition, one can uniquely determine whether a state is entangled or separable since for any separable state $\exists! \lambda_t\neq 0$. The number $r$ of nonzero $\lambda_t$'s is referred to as the Schmidt rank of the state. 

\subsection{Second quantization state space and entanglement}
In quantum field theory, the state space for a given field is given by the Fock space. This space is constructed upon single particle Hilbert space $\mathcal{H}$  as the direct sum of the symmetrized or antisymmetrized tensor product of arbitrarily large number of copies of $\mathcal{H}$:
\begin{equation}\label{eq:Fock_cons}
    \mathcal{F}(\mathcal{H})_{+(-)} = \bigoplus_{n=0}^{\infty} \mathcal S_{+(-)}(\mathcal H^{\otimes n}),
\end{equation}
where $\mathcal S_{+}$ stands for symmetrization and $\mathcal S_{-}$ for antisymmetrization. The operation $\mathcal S_q$ accounts for the indistinguishability of particles of given spiecies with $q=+$ for bosons and $q=-$ for fermions. The infinite direct sum allows for a description of an arbitrary number of particles. Note that this space in fact does not contain states with an infinite number of particles, since all well-defined states have always zero probability to find an infinite number of quanta.  In this space, one introduces the creation $\hat a^\dagger_i$ and annihilation $\hat a_i$ operators, which add or remove a single particle in a mode $a_i$. These operators fulfill cannonical commutation or anticommutation relations for bosons and fermions respectively:
\begin{align}
    &[\hat a_i,\hat a^\dagger_j]=\delta_{i,j},\,\, [\hat a_i,\hat a_j]=[\hat a^\dagger_i,\hat a^\dagger_j]=0,\\ 
        &\{\hat a_i,\hat a^\dagger_j\}=\delta_{i,j},\,\, \{\hat a_i,\hat a_j\}=\{\hat a^\dagger_i,\hat a^\dagger_j\}=0.
\end{align}
Note that modes correspond  to the choice of orthonormal basis vectors in single particle space $\mathcal{H}$, and thus the construction of the creation and annihilation operators is not unique. The canonical basis in $\mathcal{F}(\mathcal{H})_q$ is the Fock basis (occupation-number basis) and it can be always constructed from the vacuum state $\ket{\Omega}$,  defined by the relation $\hat a_i\ket{\Omega}=0$, which contains zero quanta, using creation operators $\hat a^\dagger_i$:
\begin{equation}
    \ket{n_1,n_2,...}=\frac{(\hat a^\dagger_1)^{n_1}(\hat a^\dagger_2)^{n_2}...}{\sqrt{n_1!n_2!...}}\ket{\Omega},
\end{equation}
where $\sum_i n_i<\infty$ for bosons, and for fermions $n_i\in\{0,1\}$. Note that the Fock space is separable, as the countable basis of $\mathcal H$ induces a countable number of modes, and there is only a finite number of quanta in those modes.

Different types of entanglement were considered in the Fock space \cite{Wiseman03, Demkowicz15,BENATTI20201}. One of them is the concept of entanglement of indistinguishable particles emerging from the symmetrization procedure in Fock space construction. However, let us comment that this type of entanglement is not directly accessible by nature and rather seems to be a mathematical artifact of the construction of  the Fock space.  This is because one cannot build  local observables without some effective distinguishability. As an example, consider the following state of two indistinguishable bosons in two modes 
\begin{equation}
    \ket{1,1}=\frac{1}{\sqrt{2}}\left(\ket{a_1}_1\ket{a_2}_2+\ket{a_2}_1\ket{a_1}_2\right),
\end{equation}
where kets indices mark the corresponding single particle Hilbert spaces. However, those indices are not accessible as the environment has to couple to all single particle Hilbert spaces in the same way due to indistinguishability. This, however, does not preclude the possibility of utilizing indistinguishability of particles to obtain conditional multipartite interference effects, see e.g.  \cite{Blasiak19, Blasiak21, Blasiak22}, which by nature do not demand addressing of inaccessible particle labels.

The entanglement of quantum fields which can be directly observed is the mode entanglement. In this approach to entanglement, the effective distinguishability is obtained due to the orthogonality of modes, which allows for probing their properties separately by construction of mode-local observables.  Let us divide modes emerging from some single particle Hilbert space $\mathcal{H_{AB}}$ into two families $\{a_i\},\, \{b_l\}$ (where $i$ and $l$ are countable indices). A state $\ket{\psi_{a,b}}$ in the Fock space $\mathcal{F}(\mathcal{H_{AB}})_q$ is separable iff:
\begin{equation}
    \ket{\psi_{a,b}}=f(a_i^\dagger)g(b_{l}^\dagger)\ket{\Omega},
\end{equation}
where $f(a_i^\dagger),\,g(b_{l}^\dagger)$ are some polynomials of the creation operators in modes $\{a_i\},\, \{b_l\}$ respectively. 
\section{Schmidt decomposition in Fock space}
Let us start by an explicit construction of the Schmidt decomposition in a Fock space. Note that construction of the Fock space \eqref{eq:Fock_cons} has no natural form of a tensor product of two Hilbert spaces $\mathcal{H}_1\otimes\mathcal{H}_2$ which is used in a standard Schmidt decomposition. Therefore, as a first step, we prove that the Fock space $\mathcal{F}(\mathcal{H_{AB}})_q$ is isomorphic to the tensor product of two abstract separable Hilbert spaces $\mathcal{H_A}\otimes\mathcal{H_B}$. Note that in the following, the type of field determined by $q$ will not play any role, and the result will be valid for both. 

Let us denote by $\{\vec n^j\}_{j=1}^\infty$ and $\{\vec m^{j'}\}_{j'=1}^\infty$ sets of all vectors of the form $\vec n^j=(n^j_1,n^j_2,...)$ and $\vec m^{j'}=(m^{j'}_1,m^{j'}_2,...)$ where $\sum^{\overline{\{a_i\}}}_{i=1} n^j_i<\infty, \,\sum^{\overline{\{b_{l}\}}}_{l=1}m^{j'}_{l}<\infty$. 
These two sets describe all possible distributions of a finite number of particles among two groups of modes $\{a_i\}$ and $\{b_l\}$, which can contain in principle an infinite (though countable) number of modes.
Note that an arbitrary element of an occupation-number basis of  $\mathcal{F}(\mathcal{H_{AB}})$ can be represented by two integer indices $j,j'$ as follows:
\begin{eqnarray}
\ket{j,j'}&=&\ket{\vec n^j; \vec m^{j'}}\nonumber\\
&=&\frac{1}{\mathcal{N}_a\mathcal{N}_b}\prod_{i}(\hat a_i^\dagger)^{n_i^j}\prod_l(\hat b_l^\dagger)^{m^{j'}_l}\ket{\Omega}\nonumber\\
&=&f(a_i^\dagger,j)g(b_l^\dagger,j')\ket{\Omega},
\end{eqnarray}
in which $\mathcal{N}_{a/b}$ is a normalization factor, and $f(a_i^\dagger,j), \,g(b_l^\dagger,j')$ denote suitable monomials of creation operators in modes $a_i,\, b_l$ respectively.
Under such constraints $\{\ket{j,j'}\}_{jj'}$ is a countable orthogonal basis of  $\mathcal{F(\mathcal{H_{AB}})}$ with natural inner product:
\begin{equation}
    \label{injj}
    \braket{jj'|kk'}=\delta_{\vec n^j,\vec n^k}\delta_{\vec m^{j'},\vec m^{k'}}=\delta_{j,k}\delta_{j',k'},
\end{equation}
in which the deltas are equal to one iff the two vectors are equal and zero otherwise. 

Since the monomials $f(a_i^\dagger,j), \,g(b_l^\dagger,j')$ are uniquely defined via indices $j,j'$, we can construct two Hilbert spaces $\mathcal{H}_{A},\,\mathcal{H}_{B}$  over scalars  $\mathbb{C}$ as spanned by abstract orthogonal basis vectors $\tilde f(j),\,\tilde g(j')$ respectively, with scalar products specified by:
\begin{eqnarray}
    \braket{\tilde f(j)|\tilde f(k)}&=&\delta_{j,k}\nonumber\\
    \braket{\tilde g(j')|\tilde g(k')}&=&\delta_{j',k'}.
\end{eqnarray}
The vectors  $\tilde f(j),\,\tilde g(j')$ are in one-to-one correspondence with $f(a_i^\dagger,j), \,g(b_l^\dagger,j')$, however, they \textit{forget} about the mode structure specified by $a_i,b_l$.
Now arbitrary vector in  $\mathcal{H}_{A}$ has a form
$w=\sum_j w_j\tilde{f}(j)$, and a scalar product between arbitrary two vectors $w$ and $z$ reads:
\begin{equation}
    \braket{w|z}_{\mathcal{H}_A}=\sum_j (w_j)^*z_j,
\end{equation}
and analogous relations hold for vectors in $\mathcal{H}_{B}$. Note that, since the set of indices $\{j,j'\}$ is countable, Hilbert spaces $\mathcal{H}_{A(B)}$ are in fact separable.

Now we define a linear map $\mathcal I:\mathcal{F}(\mathcal{H_{AB}})\rightarrow \mathcal{H_A}\otimes\mathcal{H_B}$ by specifying its action on basis elements of $\mathcal{F(\mathcal{H_{AB}})}$:
\begin{equation}\label{eq:isomorphism}
    \mathcal I(\ket{j,j'})=\mathcal I\left( f(a_i^\dagger,j)g(b_l^\dagger,j')\ket{\Omega}\right):=\tilde f(j)\otimes \tilde g(j'),
\end{equation}
in which the tensor product in the definition refers to a standard algebraic tensor product between $\mathcal H_{\mathcal A}$ and $\mathcal H_{\mathcal B}$.
The map $\mathcal I$ is clearly bijective. It also preserves scalar product:
\begin{eqnarray}
     &\braket{kk'|jj'}=\delta_{k,j}\delta_{{k'},{j'}}=&\nonumber\\
     &\bra{\Omega}g(b_l^\dagger,{k'})^\dagger f(a_i^\dagger,k)^\dagger  f(a_i^\dagger,j)g(b_l^\dagger,{j'})\ket{\Omega}&\nonumber\\
&\xrightarrow[]{\mathcal I()} \braket{\tilde f(k)\otimes \tilde g(k')|\tilde f(j)\otimes \tilde g(j')}_{\mathcal{H}_{\mathcal A}\otimes\mathcal{H}_{\mathcal B}}=&\nonumber \\  &\braket{\tilde f(k)|\tilde f(j)}_{\mathcal{H}_{A}}\braket{\tilde g(k')|\tilde g(j')}_{\mathcal{H}_{B}}=\delta_{k,j}\delta_{{k'},{j'}}.&\nonumber\\
\end{eqnarray}
Therefore, map $\mathcal I$ is an isomorphism, and $\mathcal{F}(\mathcal{H_{AB}})$ and $ \mathcal{H_A}\otimes\mathcal{H_B}$ are isomorphic. Note that this is not a \textit{canonical} isomorphism, since it is basis-dependent, similarly to a Choi-Jamiolkowski isomorphism \cite{Choi75, Jamiolkowski72}.

Since by construction both $\mathcal{H_A}$ and $\mathcal{H_B}$ are separable Hilbert spaces, any vector $\phi\in\mathcal{H_A}\otimes\mathcal{H_B}$ possesses a Schmidt decomposition:
\begin{equation}
\label{schmidtHAB}
    \phi=\sum_t \lambda_t \tilde F(t)\otimes \tilde G(t),
\end{equation}
 with $\tilde F(t),\,\tilde G(t)$ being some  orthonormal basis vectors in spaces $\mathcal{H}_{A(B)}$.

Now, let us consider an arbitrary state $\ket{\psi}$ in $\mathcal{F}(\mathcal{H_{AB}})$ and find its Schmidt decomposition. By using isomorphism $\mathcal I$ one gets:
\begin{multline}
    \ket{\psi}=\mathcal I^{-1}( \mathcal I (\ket{\psi}))=\mathcal I^{-1}\left(\sum_t \lambda_t \tilde F(t)\otimes \tilde G(t) \right)=\\ \sum_t \lambda_t \mathcal I^{-1}\left(\tilde F(t)\otimes \tilde G(t) \right),
\end{multline}
in which we used Schmidt decomposition \eqref{schmidtHAB} for the vector $\mathcal I (\ket{\psi})$. Since $\tilde F(t)$ and $\tilde G(t)$ are clearly linear combinations of original basis vectors $\tilde F(t)=\sum_j q_j^t\tilde f(j)$, $\tilde G(t)=\sum_{j'} p_{j'}^t\tilde g(j')$,  
 we have:
\begin{multline}
    \mathcal I^{-1}\left(\tilde F(t)\otimes \tilde G(t) \right)=\\
  \left[\sum_j q_j^t f(a_i^\dagger,j)\right] \left[\sum_{j'} p_{j'}^t g(b_l^\dagger,{j'})\right]\ket{\Omega}=\\
  F(a_i^\dagger,t)G(b_l^\dagger,t)\ket{\Omega},
\end{multline}
where $F(a_i^\dagger,t), \,G(b_l^\dagger,t)$ are polynomials of creation operators in corresponding modes.

Finally we obtain:
\begin{equation}
    \ket{\psi}=\sum_t \lambda_t F(a_i^\dagger,t)G(b_l^\dagger,t)\ket{\Omega}.\label{eq:fock_schmidt}
\end{equation}
As inverse isomorphism $\mathcal I^{-1}$ preserves orthogonality, all terms in  \eqref{eq:fock_schmidt} are orthogonal to each other and what is more locally orthogonal:
\begin{equation}
   \bra{\Omega}F(a_i^\dagger,t)^\dagger F(a_i^\dagger,t')\ket{\Omega}= \bra{\Omega}G(b_l^\dagger,t)^\dagger G(b_l^\dagger,t')\ket{\Omega}=\delta_{t,t'}.
\end{equation}
This provides a generalization of Schmidt decomposition for an arbitrary vector in a Fock space, built up from a single-particle space with two separated groups of modes.

\section{Gisin's theorem for quantum fields}
Let us now prove the central result of the work which states that for any pure entangled state there exists a Bell inequality which is violated. The state of the field $\ket{\psi}$ is separable with respect to the partition of modes into families $\{a_i\},\, \{b_l\}$  if it can be put as $\ket{\psi}=F(a_i^\dagger)G(b_l^\dagger)\ket{\Omega}$ which corresponds to $\exists! \alpha_t\neq 0$ in Schmidt decomposition \eqref{eq:fock_schmidt} and otherwise state is entangled. Since any state admits Schmidt decomposition  \eqref{eq:fock_schmidt},  a state is entangled iff decomposition has at least two non-zero terms i.e., it has a form:
\begin{multline}
\label{fieldSchmidt}
    \ket{\psi_{ent}}= \\\lambda_1 F(a_i^\dagger,1)G(b_l^\dagger,1)\ket{\Omega}+\lambda_2 F(a_i^
\dagger,2)G(b_l^\dagger,2)\ket{\Omega}\\+ \lambda_R\ket{R}= \sqrt{|\lambda_1|^2+|\lambda_1|^2}\ket{\phi_{12}}+ \lambda_R\ket{R},
\end{multline}
where $\ket{R}$ is some state emerging from the terms for $t>2$ which is orthogonal to the first two terms and $\lambda_R$ is the amplitude corresponding to this state (it can be zero). Now, assume that the set $\mathcal{S}:=\{F(a_i^\dagger,t)G(b_l^\dagger,t)\ket{\Omega}\}_t$ is complete (spans the entire Fock space  $\mathcal{F}(\mathcal{H_{AB}})$). This holds without loss of generality, as if the decomposition of the state $\ket{\psi}$ does not contain terms corresponding to the full complete set, one can extend the set $\mathcal{S}$ to a complete set by choosing an orthogonal basis of the Schmidt form in the orthogonal complement of this set. In such a case, the basis elements of the extended part of the set $\mathcal{S}$ are included in the Schmidt decomposition of a state $\ket{\psi}$ simply with $\lambda_t=0$. Consider now a subspace of a space of all well-defined operators on the Fock space  $\mathcal{F}(\mathcal{H_{AB}})$ spanned by the following operators:
\begin{align} \label{eq:sub_p}
P_{ab}^{k,k',n,n'}=F(a_i^\dagger,k)G(b_l^\dagger,n)\ket{\Omega}\bra{\Omega}G(b_l^\dagger,n')^\dagger F(a_i^\dagger,k')^\dagger ,
\end{align}
where $k,k',n,n'\in\{1,2\}$. Local operators which represent observables local with respect to the parties can be written as:
\begin{align} \label{eq:sub_p_loc}
    P_{a}^{k,k'}&=\sum_t F(a_i^\dagger,k)G(b_l^\dagger,t)\ket{\Omega}\bra{\Omega}G(b_l^\dagger,t)^\dagger F(a_i^\dagger,k')^\dagger ,\nonumber\\
     P_{b}^{k,k'}&=\sum_t F(a_i^\dagger,t)G(b_l^\dagger,k)\ket{\Omega}\bra{\Omega}G(b_l^\dagger,k')^\dagger F(a_i^\dagger,t)^\dagger.
\end{align}
Clearly expectation value of any operator  $\hat O$ from the subspace spanned by operators \eqref{eq:sub_p}, \eqref{eq:sub_p_loc} calculated on the state $\ket{\psi_{ent}}$ consists of two terms:
\begin{multline}
\label{averageOnSchmidt}
    \braket{\hat O}_{\psi_{ent}}= (|\lambda_1|^2+|\lambda_2|^2)\braket{\hat O}_{\phi_{12}}+|\lambda_R|^2\braket{\hat O}_{R}=\\ (|\lambda_1|^2+|\lambda_2|^2)\braket{\hat O}_{\phi_{12}},
\end{multline}
where the second term in the first row is equal to zero as the state $\ket{R}\bra{R}$ is in the subspace orthogonal to all operators \eqref{eq:sub_p}. 

 Clearly $\ket{\phi_{12}}$ is entangled in the considered $2\times 2$ dimensional subspace, and therefore from the original Gisin's theorem there exists a Clauser-Horne-Shimony-Holt (CHSH) inequality, which is violated by such a state on the positive side.
 Following \cite{Gisin_PLA} we can directly construct such a CHSH inequality and the corresponding observables and settings leading to a violation. The CHSH inequality from Gisin's original work has the following conditional form:

 \begin{eqnarray}
 \label{CHSH_orig}
     E(\vec \alpha, \vec \beta)-E(\vec \alpha, \vec \beta')+E(\vec \alpha', \vec \beta)+E(\vec \alpha', \vec \beta')\leq 2,&&\nonumber\\
     \textrm{ if } E(\vec \alpha, \vec \beta)\geq E(\vec \alpha, \vec \beta')&&\nonumber\\
     E(\vec \alpha, \vec \beta')-E(\vec \alpha, \vec \beta)+E(\vec \alpha', \vec \beta)+E(\vec \alpha', \vec \beta')\leq 2,&&\nonumber\\
     \textrm{ if } E(\vec \alpha, \vec \beta)< E(\vec \alpha, \vec \beta'),&&
 \end{eqnarray}
 where the correlation function for two-qubit system is defined as $E(\vec \alpha, \vec \beta)=\langle \vec \alpha \cdot \vec\sigma\otimes\vec \beta\cdot\vec\sigma\rangle$. The settings are parametrized by vectors:
 \begin{eqnarray}
     \vec \alpha&=&(\sin \alpha,0, \cos \alpha)\nonumber\\
       \vec \beta&=&(\sin \beta,0, \cos \beta)\nonumber\\
       \vec \alpha'&=&(\sin \alpha',0, \cos \alpha')\nonumber\\
         \vec \beta'&=&(\sin \beta',0, \cos \beta').
 \end{eqnarray}
 Assuming the two-qubit state in a specific Schmidt form:
 \begin{equation}
 \label{schmidtQubit}
     \ket{\psi}=\lambda_1\ket{01}+\lambda_2\ket{10},
 \end{equation}
 the optimal settings are specified by the following angles:
 \begin{eqnarray}
     &&\alpha=0,\nonumber\\
     &&\alpha'=-\operatorname{sign}(\lambda_1\lambda_2)\frac{\pi}{2}\nonumber\\
     &&\cos\beta=-\cos\beta'=\left(1+4|\lambda_1\lambda_2|\right)^{-\frac{1}{2}}.
 \end{eqnarray}
 In order to apply the CHSH inequality  \eqref{CHSH_orig} to the state $\ket{\phi_{12}}$ one has to translate qubit observables:  $$\vec \alpha \cdot \vec\sigma\otimes\vec \beta\cdot\vec\sigma$$  to field operators spanned by the operators \eqref{eq:sub_p_loc} local with respect to the parties. Since the two effective qubit subspaces in $\ket{\phi_{12}}$ are spanned by operators $\{F(a_i^\dagger,1), F(a_i^\dagger,2)\}$ and respectively  $\{G(b_l^\dagger,1), G(b_l^\dagger,2)\}$, the field operators representing qubit Pauli observables have the following form:
 \begin{eqnarray}
 \label{fieldSigmas}
     \Sigma_x(a_i)&=&P_a^{12}+P_a^{21},\nonumber\\
     \Sigma_y(a_i)&=&-i(P_a^{12}-P_a^{21}),\nonumber\\
     \Sigma_z(a_i)&=&P_a^{11}-P_a^{22},\nonumber\\
     \Sigma_x(b_l)&=&P_b^{12}+P_b^{21},\nonumber\\
     \Sigma_y(b_l)&=&-i(P_b^{21}-P_b^{12}),\nonumber\\
     \Sigma_z(b_l)&=&P_b^{22}-P_b^{11},
 \end{eqnarray}
where for the operators on modes $\{b_l\}$ a flip of basis elements is taken into account in accordance with \eqref{schmidtQubit}
(namely for the second effective qubit $G(b_l^\dagger, 1)$ plays the role of $\ket{1}$ and $G(b_l^\dagger,2)$ of $\ket{0}$).  Finally, an arbitrary qubit observable is mapped into the following field observable:
\begin{equation}
    \vec \alpha \cdot \vec\sigma\otimes\vec \beta\cdot\vec\sigma\mapsto\left( \vec \alpha \cdot \vec\Sigma(a_i)\right)\left( \vec \beta \cdot \vec\Sigma(b_i)\right), 
\end{equation}
in which $\vec\Sigma(a_i)=(\Sigma_x(a_i), \Sigma_y(a_i), \Sigma_z(a_i))$ and analogously for $\vec\Sigma(b_l)$.  Finally, the CHSH inequality for fields is built using the following correlation function:
\begin{equation}
    \mathcal E(\vec \alpha, \vec \beta)_{\phi_{12}}=\left\langle\left( \vec \alpha \cdot \vec\Sigma(a_i)\right)\left( \vec \beta \cdot \vec\Sigma(b_i)\right)\right\rangle_{\phi_{12}},
\end{equation}
 in which the caligraphic notation on the left-hand-side indicates that we deal with the correlation function for fields not for qubits. By construction, one of the following CHSH inequalities is always violated for the state $\ket{\phi_{12}}$:
 \begin{eqnarray}
 \label{CHSH}
     \mathcal E(\vec \alpha, \vec \beta)_{\phi_{12}}- \mathcal E(\vec \alpha, \vec \beta')_{\phi_{12}}+ \mathcal E(\vec \alpha', \vec \beta)_{\phi_{12}}+ \mathcal E(\vec \alpha', \vec \beta')_{\phi_{12}}\leq 2,&&\nonumber\\
     \textrm{ if }  \mathcal E(\vec \alpha, \vec \beta)_{\phi_{12}}\geq  \mathcal E(\vec \alpha, \vec \beta')_{\phi_{12}}&&\nonumber\\
     \mathcal E(\vec \alpha, \vec \beta')_{\phi_{12}}- \mathcal E(\vec \alpha, \vec \beta)_{\phi_{12}}+ \mathcal E(\vec \alpha', \vec \beta)_{\phi_{12}}+ \mathcal E(\vec \alpha', \vec \beta')_{\phi_{12}}\leq 2,&&\nonumber\\
     \textrm{ if }  \mathcal E(\vec \alpha, \vec \beta)_{\phi_{12}}<  \mathcal E(\vec \alpha, \vec \beta')_{\phi_{12}}.&&\nonumber\\
 \end{eqnarray}
  Due to the factor $ (|\lambda_1|^2+|\lambda_2|^2)$ in the formula \eqref{averageOnSchmidt}, the above inequalities \textit{need not be violated on the entire initial state } $\ket{\psi_{ent}}$. However, we can easily prove that the corresponding Clauser-Horne inequalities \textit{are violated} on  $\ket{\psi_{ent}}$ whenever the CHSH inequalities are violated solely on $\ket{\phi_{12}}$.
  Let us start from defining these new inequalities. In $2\times 2$ dimensional spaces a positive violation of  the CHSH inequality is equivalent to a positive violation of a Clauser-Horne (CH) inequality  (see e.g. \cite{CH_eq}, section 2.5.2).
  The correspondence between the two inequalities is due to the following relation between correlation functions and probabilities:
  \begin{equation}
      E(\vec \alpha,\vec \beta)=4p(0,0|\vec \alpha, \vec \beta)-2p(0|\vec \alpha)-2p(0|\vec \beta)+1, 
  \end{equation}
  which holds whenever the outcomes of the measured observables are dichotomic. In the above we have chosen outcomes labeled by $(0,0)$, however any choice of the outcomes is valid, what is important is that they are fixed in the above relation.  In our scenario the field observables \eqref{fieldSigmas} are dichotomic when restricted to the subspace spanned by first two Schmidt bases elements in \eqref{fieldSchmidt}, therefore, on this subspace the mapping between CHSH and CH inequality holds exactly. The CH inequalities corresponding to the CHSH inequalities \eqref{CHSH} have the following form:
      \begin{eqnarray}
      \label{fieldCH}
          p(0,0|\vec \alpha, \vec \beta)+p(0,0|\vec \alpha', \vec \beta)+p(0,0|\vec \alpha', \vec \beta')-p(0,0|\vec \alpha, \vec \beta')&&\nonumber\\
          -p(0|\vec \alpha')-p(0|\vec \beta)\leq 0&&\nonumber\\
          p(0,0|\vec \alpha, \vec \beta')+p(0,0|\vec \alpha', \vec \beta)+p(0,0|\vec \alpha', \vec \beta')-p(0,0|\vec \alpha, \vec \beta)&&\nonumber\\
          -p(0|\vec \alpha')-p(0|\vec \beta')\leq 0,&&\nonumber\\
      \end{eqnarray}
  in which the probabilities are specified as follows:
  \begin{eqnarray}
      \label{probCHop}
      &p(0,0|\vec \alpha, \vec \beta)=\left\langle \frac{1}{4}(\id_A+\vec \alpha\cdot\vec\Sigma(a_i))(\id_B-\vec \beta\cdot\vec\Sigma(b_l))\right\rangle_{\phi_{12}}&\nonumber\\
      &p(0|\vec \alpha)=\left\langle\frac{1}{2}(\id_A+\vec \alpha\cdot\vec\Sigma(a_i))\right\rangle_{\phi_{12}}&\nonumber\\
      &p(0|\vec \beta)=\left\langle\frac{1}{2}(\id_B-\vec \beta\cdot\vec\Sigma(b_l))\right\rangle_{\phi_{12}}.&
  \end{eqnarray}
 In the above formulas $\id_{A/B}$ denotes the following operators which act as projectors onto the local subspace of interest and identity on the second subsystem:
\begin{align}
\begin{split}
  \id_A&=P_{a}^{1,1}+P_{a}^{2,2},\\
  \id_B&=P_{b}^{1,1}+P_{b}^{2,2}.
  \end{split}
\end{align}
  Now violation of one of the CHSH inequalities \eqref{CHSH} on $\ket{\phi_{12}}$ implies violation of the corresponding CH inequality from among \eqref{fieldCH} on   $\ket{\phi_{12}}$. It suffices to show that the CH inequality is then also violated on the entire state $\ket{\psi_{ent}}$.
  Let us denote the Bell operator for a CH inequality as $\hat{CH}$. Then whenever any of the CH inequalities \eqref{fieldCH} is violated, we have for the corresponding Bell operator: $\braket{\hat{CH}}_{\phi_{12}}>0$.  This yields:
\begin{equation}
    \braket{\hat{CH}}_{\psi_{ent}}= (|\lambda_1|^2+|\lambda_1|^2)\braket{\hat{CH}}_{\phi_{12}}>0,\label{eq:ch_g}
\end{equation}
an therefore violation of the CH inequality on the state $\ket{\phi_{12}}$ implies a violation of the inequality by the state $\ket{\psi_{ent}}$. 
This ends the proof.

Note that the violation of the CHSH inequalities on the state $\ket{\phi_{12}}$ \eqref{CHSH} \textit{does not imply} violation of these inequalities on the entire state $\ket{\psi_{ent}}$, therefore the transition to the CH inequality is necessary for the proof to hold. The two inequalities are \textit{not equivalent} on the entire state space, since then the measurements defined by operators \eqref{fieldSigmas} are no longer dichotomic.
One has to be very careful when trying to prove CHSH inequality violation on the entire state space based on its violation on the subspace, since this can lead to false conclusions on Bell nonclassicality, see e.g. \cite{DasDunnComm}.

\subsection{Two distinguishable fields}
One can also consider the entanglement of two or any finite number of species of fields. For simplicity, let us consider the case of two fields as the other cases follow analogously.  In such a case the state space is given by the tensor product of two Fock spaces  $\mathcal{F}(\mathcal{H}_{AB})_q\otimes\mathcal{F'(\mathcal{H}_{A'B'})}_{q'}$ where we already assumed partition of modes into families $\{a_i\},\, \{b_l\}$ and $\{a'_i\},\, \{b'_l\}$. Note that one of those families of modes per primed or unprimed pair could be empty. Once again, one can represent the occupation number basis vectors for the first field as  $\vec n^j, \vec m^{j'}$ and $\vec{(n')}^j,\vec{(m')}^{j'}$ for the second using suitable monomials of creation operators in modes corresponding to the considered partition:
\begin{multline}
    \ket{\vec n^j,\vec{(n')}^j; \vec m^{j'},\vec{(m')}^{j'}}\\=f(a_i^\dagger,(a_i')^\dagger,j)g(b_{l}^\dagger,(b_{l}')^\dagger,j')\ket{\Omega},
\end{multline}
where indices $j,j'$  are again countable and $\ket{\Omega}$ stands for tensor product of vacuum states of both fields. We use here the fact that creation operators for distinguishable fields commute, and therefore one can group them as it is presented on the right-hand side.
 Now one can see that isomorphism \eqref{eq:isomorphism}  can be trivially generalized to this case. This results in isomorphism between space $\mathcal{F}(\mathcal{H}_{AB})_q\otimes\mathcal{F'(\mathcal{H}_{A'B'})}_{q'}$  and separable space $\mathcal{H}_{A,A'}\otimes\mathcal{H}_{B,B'}$ where $\mathcal{H}_{A,A'(B,B')}$ are counterparts to  $\mathcal{H}_{A,(B)}$. Once again, one can apply the Schmidt decomposition on the state mapped to $\mathcal{H}_{A,A'}\otimes\mathcal{H}_{B,B'}$ in order to obtain the decomposed state in the tensor product of Fock spaces. The resulting general form of the Schmidt decomposition of the state $\ket{\psi}\in\mathcal{F}(\mathcal{H}_{AB})_q\otimes\mathcal{F'(\mathcal{H}_{A'B'})}_{q'}$ is the following:
\begin{equation}
    \ket{\psi}=\sum_t \lambda_t F(a_i^\dagger,(a'_i)^\dagger,t)G(b_l^\dagger,(b'_l)^\dagger,t)\ket{\Omega},
\end{equation}
where $F(a_i^\dagger,(a'_i)^\dagger,t),\,G(b_l^\dagger,(b'_l)^\dagger,t)$ are polynomials of creation operators of respective modes. Upon this for any entangled pure state one can build a CH inequality, which would be violated in full analogy to the construction ending at formula \eqref{eq:ch_g}.
\section{Non-separable Hilbert spaces}

The issue of whether non-separable Hilbert spaces are useful for Quantum Field Theory is subtle and controversial. On the one hand, a separable Hilbert space is sufficient to describe scattering processes (see, e.g. \cite{PCT89}, Chapter 2.6). On the other hand, for some applications it is convenient to introduce \textit{big} Hilbert space $\mathcal H' = \bigoplus_{n=0}^{\infty} \mathcal H^{\otimes n}$, where $\mathcal H^{\otimes n}$ for $n\rightarrow\infty$ is treated as an \textit{infinite tensor product}, which describes the system of \textit{actually infinite} number of particles. Such \textit{big} Hilbert space is non-separable and contains as its  subspaces all Fock spaces  emerging from all different choices of basis solutions for field equations   (these Fock spaces are defined in a similar  way, but with $\mathcal H^{\otimes n}$ interpreted as representing only \textit{potentially infinite} number of particles). This \textit{big}  Hilbert space $\mathcal H'$ is, in fact, built from those Fock spaces. The creation and annihilation operators corresponding to these Fock spaces are then related by non-unitary transformations (for example, by Bogoliubov transformations) leading to appearance of an effective description in terms of quasi-particles \cite{Kitamura12}. Then the infinite particle states are effectively hidden in different "vacuum" states of  the corresponding Fock spaces. \textit{Big} non-separable Hilbert space is useful, for example, in statistical mechanics for description of phase transitions: taking the thermodynamic limit, in which the density of a system remains constant, while its spatial dimensions tend to infinity, demands considering actually infinite number of particles in the system. 

Therefore, in principle, one can consider the \textit{ big} multimode Hilbert space $\mathcal H'$, with modes grouped into two disjoint subsets $\mathcal A$ and $\mathcal B$, which define the local structure for a Bell-type reasoning. This space if formally spanned by number states $\{\ket{\vec n_{\nu}}_{\mathcal A}\ket{\vec m_{\mu}}_{\mathcal B}\}$, in which vectors $\vec n_{\nu}$ and $\vec m_{\mu}$ represent infinite sequences of integers without cut-off conditions used in defining the Fock space. Such sets of vectors are certainly uncountable. Note that the cardinality of this set of basis vectors is $\aleph_0^{\aleph_0}$ which is the same as saying that its cardinality is of the continuum $\mathfrak{c}$ and therefore it can be indexed symbolically by real indices $\nu, \mu$. In order to justify in what sense Gisin's theorem generalizes to such a case, we have to recall a very important result in mathematical physics, namely generalization of Gleason's theorem for non-separable Hilbert spaces. Recall that in the case of separable Hilbert spaces, Gleason's theorem states that any assignment $m: \{\hat\Pi_i\}\mapsto [0,1]$ of a probability distribution to a family of projectors onto closed subspaces of a Hilbert space of dimension greater than two is represented by $m(\hat\Pi_i)=\Tr(\rho_m\hat\Pi_i)$, in which $\rho_m$ is a normalized positive-semidefinite matrix, typically called a density matrix. This theorem has been generalized to the case of a non-separable Hilbert space \cite{Eilers75}. This generalization states that if the cardinality of a set of basis vectors in a given Hilbert space is given by the Ulam number and under the assumption of the continuum hypothesis, any assignment of a probability distribution  $m: \{\hat\Pi_{\eta}\}\mapsto [0,1]$ to a (possibly uncountable) family of projectors onto closed subspaces of the Hilbert space is represented by $m(\hat\Pi_{\eta})=\Tr(\rho_m\hat\Pi_{\mathcal K_m}\hat\Pi_{\eta})$, in which $\hat\Pi_{\mathcal K_m}$ is a projector onto a separable subspace $\mathcal K_m$ of the entire Hilbert space and $\rho_m$ is a state on the separable space $\mathcal K_m$. Therefore, one can say that any state on a non-separable Hilbert space is in fact represented by a state on a separable subspace. In the context of the \textit{big} Hilbert space introduced in the previous paragraph, as continuum $\mathfrak{c}$ is an Ulam number, this theorem implies that any state on the \textit{big} space is represented by a state belonging to some Fock space or at most finite combinations of them. Since our generalization of Gisin's theorem works for any state on any Fock space, it effectively works for any state defined on the \textit{big} Hilbert space. In Appendix \ref{app:A} the Schmidt decomposition of the state in $\mathcal H'$ in terms of extended occupation number basis is presented, which then allows for analogous direct construction of a CH inequality that is always violated for an entangled state of the field modes.

\section{Examples}
Let us consider a few examples of Schmidt decompositions of entangled states with undefined particle numbers. The first example is of a particular importance as for this state one cannot reveal  its entanglement by projecting it into specific particle numbers and trying to map the problem to the language of first quantization. Therefore, this example in particular was not covered by previous generalizations of Gisin's theorem. This example is the two-mode squeezed vacuum state which in its Schmidt form can be written as:
\begin{equation}
\ket{\Psi}=\sum_{n=0}^\infty\lambda_n (\hat a_1^\dagger)^{n}(\hat b_1^\dagger)^{n}\ket{\Omega}=
\sum_{n=0}^\infty\lambda_n\ket{n_{a_1};n_{b_1}}.\label{eq:bsv2}
\end{equation}
 Now one can choose the following partition for investigating the CH inequality violation:
\begin{multline}
   \ket{\Psi}=N_{\lambda_0\lambda_1}\left[\frac{1}{\sqrt{|\lambda_0|^2+|\lambda_1|^2}}\left(\lambda_0\ket{0;0}+\lambda_1\ket{1;1}\right)\right]
\\+\lambda_R\ket{R},
\end{multline}
in which $N_{\lambda_0\lambda_1}=\sqrt{|\lambda_0|^2+|\lambda_1|^2}$.
Here, the construction of the CH operator based  on the operators  \eqref{fieldSigmas}  requires the ability to implement measurements which perform projections onto states of the form  $\cos\theta\ket{0}+e^{i\phi}\sin\theta\ket{1}$. It is, however, not known how to perform such a projection experimentally. One can find a violation of some Bell inequality for states from subspace in which the CH operator is constructed using homodyne measurements \cite{SinglePhotonGPY}. However, such measurements cannot project the state $\ket{\Psi}$ onto this subspace. Thus, testing CH inequalities based on Schmidt decomposition is not always feasible experimentally, and this problem is described in more detail in the next section. Still, this is only a technical limitation, and fundamentally this state does not admit local hidden variable model.

However, there exist states with an undefined photon number for which one can experimentally test the CH inequality violation constructed as  in\eqref{eq:ch_g}. As an example we consider the $2\times 2$ bright squeezed vacuum state which  can be put into the following Schmidt form :
\begin{multline}
\ket{\psi_-}=
\frac{1}{\cosh^2(\Gamma)}\sum_{n=0}^\infty\tanh^n(\Gamma)\\\sum_{m=0}^n(-1)^m\ket{(n-m)_{a_1},m_{a_2};m_{b_1},(n-m)_{b_2}},\label{eq:bsv}
\end{multline}
where the parameter $\Gamma$ denotes the amplification gain. Consider the following partition of the state:
\begin{multline}\label{eq:bsv_ex}
    \ket{\psi_-}=\frac{\sqrt{2}\tanh(\Gamma)}{\cosh^2(\Gamma)}\left[\frac{1}{\sqrt{2}}(\ket{1,0;0,1}-\ket{0,1;1,0})\right]\\+ \alpha_R \ket{R}
\end{multline}
where the first term is a typical double-rail-encoded singlet. The measurements  corresponding to the operator subspace \eqref{eq:sub_p_loc} are simply given by Stokes operators or Sign Stokes operators \cite{ZUKUPRA, schlichtholz2021simplified} restricted to single particle subspace and act on it like Pauli operators. Such operators  can be implemented using photon number resolving detectors.
However, taking another partition of the state \eqref{eq:bsv} including states with more than one photon, one would need to, for example, implement generalized Pauli operators for quantum fields \cite{Schlichtholz_2022}, the experimental realization of which is unclear.

As a final example we consider  the BGHZ state \cite{BGHZ} which has the following Schmidt form: 
\begin{align}
\begin{split}
\ket{BGHZ}=&\sum_{k=0}^{\infty}\sum_{m=0}^kC_{k-m}(\Gamma)C_m(\Gamma)\\
&\times( \hat a_1^{\dagger} \hat a_2^{\dagger} \hat b_1^{\dagger})^{k-m}( \hat a_3^{\dagger}\hat a_4^{\dagger}\hat b_2^{\dagger})^m\ket{\Omega},
\end{split}
\label{BGHZ0} 
\end{align}
where $C_q(\Gamma)$ are coefficients dependent on amplification gain $\Gamma$. Now we can choose the relevant terms in the Schmidt decomposition for the construction of the CH inequality as follows:
\begin{multline}
    \ket{BGHZ}=\alpha_{12} \left[\frac{1}{\sqrt{2}}(\ket{1,1,0,0;1,0}+\ket{0,0,1,1;0,1})\right]\\+ \alpha_R \ket{R}.
\end{multline}
The first term is simply polarization or path GHZ state. However, the necessary projectors for  the CH inequality \eqref{fieldCH} violation have no  known experimental realization as it requires performing Bell-state measurement within passive linear optics. However, if one considers BGHZ state after single-photon subtraction among modes $a_2,a_4$ one gets:
\begin{multline}
\ket{BGHZ}_{\operatorname{s}}=\alpha_{12}' \left[\frac{1}{\sqrt{2}}(\ket{1,0,0,0;1,0}+\ket{0,0,1,0;0,1})\right]\\+ \alpha_R' \ket{R}.
\end{multline}
Due to this modification  the suitable measurements for CH inequality violation have their experimental realization again as  Stokes operators restricted to single-photon subspace.

One can see that while the construction of the CH inequality \eqref{fieldCH} is purely abstract, in some circumstances its violation could be, in principle, measured using current technology.

\section{Sufficient condition for the necessity of applying non-projective POVMs or other tricks}

As noted in the previous section, there are many entangled states of quantum fields for which projective measurements are not sufficient to reveal their non-classical behavior. The reason for that is the impossibility of experimental realization of  projections onto the states which are superpositions of states with different numbers of particles in the same mode like $(\ket{0}+\ket{1})/\sqrt{2}$. Therefore, some non-projective measurements are necessary in such circumstances. In particular in quantum optics those are realized through addition of ancillary modes with coherent states to perform homodyne measurements. One can also, at least in some cases, engineer the ancillary state for the examined state in such a way that the violation can converge to the ideal case \cite{Paterek11} (however, it is not known if the preparation of such ancillas is experimentally feasible). The Schmidt decomposition allows us to determine if such measurements are necessary. We formulate the following theorem:

\newtheorem{tw}{Theorem}
\begin{tw}
The sufficient condition for the necessity of  using ancillary resources (such as coherent states in ancillary modes or postselection) for experimental revealing of non-classicality of entangled state $\ket{\psi}$  can be formulated as follows: all projections of the state $\ket{\psi}$ into subspace given by $\sum_i n_i=n$, $\sum_{i'} m_{i'}=m$ have Schmidt rank $1$ i.e.
\begin{equation}
    \forall_{n,m} \, \frac{1}{N}\hat\Pi_{nm}\ket{\psi}=F^n(a_i^\dagger,1)G^m(b_l^\dagger,1)\ket{\Omega},\label{eq:th1}
\end{equation}
where $N$ is a normalization factor, $\Pi_{nm}$ is a projector onto the subspace considered, and $F^n(a_i^\dagger,1),\,G^m(b_l^\dagger,1)$ are polynomials of creation operators of the order $n,\,m$ respectively. 
\end{tw}
Note that one can write projector $\hat \Pi_{n,m}$ as follows:
\begin{equation}
  \hat \Pi_{n,m}=\sum_{j|\sum n^j_i=n} \sum_{j'|\sum m^{j'}_i=m} \ket{\vec n^j, \vec m^{j'}}\bra{\vec n^j, \vec m^{j'}}.
\end{equation}
This Theorem is followed by the corollary:
\newtheorem{co}{Corollary}
\begin{co}
The necessary condition for revealing non-classicality of entangled state $\ket{\psi}$ with experimentally realizable projective measurements can be formulated as follows: There exists a projection of the state $\ket{\psi}$ into the subspace given by $\sum_i n_i=n$ ,$\sum_{i'} m_{i'}=m$ which have Schmidt rank higher than $1$ i.e.
\begin{multline}
    \exists_{n,m} \exists_{\alpha_i,\alpha_j\neq0}\, \frac{1}{N}\hat\Pi_{nm}\ket{\psi}\\=\sum_t\alpha_t F^n(a_i^\dagger,t)G^m(b_l^\dagger,t)\ket{\Omega}.\label{eq:col1}
\end{multline}
\end{co}
The intuitive meaning of the above two statements is the following: whenever an \textit{entangled} state of the optical field contains (mode) entanglement between terms with the same total photon number within the two subspaces corresponding to the two families of modes $\{a_i\}$ and $\{b_l\}$, then one can detect the nonclassicality of the state with projectors realizable with passive linear optics. However, if there is no entanglement of this kind present in the state, one needs to utilize additional resources like ancillary modes or postselection in order to detect nonclassicality of the state.

Let us give the proof for the theorem; the proof for colloraly follows immediately.
Any projective measurement $\hat M$ which does not project into the superposition of states with different photon numbers always acts separately on the subspaces determined by the projectors $\hat\Pi_{nm}$. Therefore, one can write:
\begin{equation}
    \braket{\hat M}_{\psi}=\sum_{n,m}\bra{\psi}\hat\Pi_{nm}\ket{\psi}\bra{\psi_{n,m}}\hat M\ket{\psi_{n,m}},
\end{equation}
where :
\begin{equation}
    \ket{\psi_{n,m}}=    \frac{\hat\Pi_{nm}\ket{\psi}}{\sqrt{\bra{\psi}\hat\Pi_{nm}\ket{\psi}}}.\label{eq:def_psi}
\end{equation}
Let $\hat M$ be the Bell operator and let the state $\ket{\psi}$ fulfill the condition \eqref{eq:th1}. This indicates that all $\ket{\psi_{nm}}$ are separable. Then the expectation value of the Bell operator is effectively taken over a separable state:
\begin{equation}
    \hat \rho_{sep}= \sum_{nm} \bra{\psi}\hat\Pi_{nm}\ket{\psi}\ket{\psi_{n,m}}\bra{\psi_{n,m}}.
\end{equation}
Therefore, the expectation value of the Bell operator cannot violate the Bell inequality. Similarly, any entanglement or steering criterion based on projective measurement considered will not revel non-classical behavior. This ends the proof.

The Corollary 1 provides only necessary condition because while the class of observables which does not mix different photon number states is clearly at least partially realized in experiments, one does not know if all such operators can be experimentally implemented. Therefore, even if this condition is fulfilled, using ancillary modes might be necessary due to experimental but not fundamental reasons.

\subsection{Introducing ancillary modes}
When introducing ancillary modes (the state of which is separable with respect to the original one), the Schmidt form is modified only by the multiplication of all terms by a single polynomial of creation operators per party. Let us denote by $A(c_k^\dagger)$, $B(d_h^\dagger)$ the polynomials that describe the state of the ancillary modes $c_k$, $d_h$ for the party $A$ and $B$ respectively. Because states of ancillary modes are separable with respect to the state of modes $a_i,b_{l}$ we can write the total state as follows:
\begin{multline}
    \ket{\Psi}=A(c_k)\ket{0}B(d_h)\ket{0}\ket{\psi}\\=\sum_t \lambda_t A(c_k^\dagger)F(a_i^\dagger,t)G(b_l^\dagger,t)B(d_h^\dagger)\ket{\Omega}\\
    =\sum_t \lambda_t F'(a_i^\dagger,c_k^\dagger,t)G'(b_l^\dagger,d_h^\dagger,t)\ket{\Omega}.\label{eq:ancila}
\end{multline}
Clearly the states generated by polynomials $F'(a_i^\dagger,c_k^\dagger,t),\,G'(b_l^\dagger,d_h^\dagger,t)$ for all $t$ are locally orthogonal. Therefore, Eq. \eqref{eq:ancila} gives Schmidt form for total state with ancillary modes.

As an example let us consider the state of a single-photon beamsplitted by a symmetric beamsplitter with output modes $a_1,b_1$ which is maximally entangled in the photon number basis:
\begin{equation}
    \ket{\phi}=\frac{1}{\sqrt{2}}(\hat a_1^\dagger+\hat b_1^\dagger)\ket{\Omega}=\frac{1}{\sqrt{2}}(\ket{0;1}+\ket{1;0}).
\end{equation}
Clearly, this state fulfills the condition \eqref{eq:th1} and therefore the entanglement of this state cannot be reached with straightforward photon number non-mixing projective measurements. Let us introduce two ancillary modes $a_2,b_2$ (one per party) occupied by coherent states with the same amplitude $\it z$ resulting in the state in its Schmidt form:
\begin{equation}\label{eq:single_full}
   \ket{z}_{a_2} \ket{\phi}\ket{z}_{b_2}=\frac{1}{\sqrt{2}}\ket{z}_{a_2}\hat a_1^\dagger\ket{0;0}\ket{z}_{b_2}+\frac{1}{\sqrt{2}}\ket{z}_{a_2}\hat b_1^\dagger\ket{0;0}\ket{z}_{b_2}.
\end{equation}
Now consider the projection of this state into the subspace given by $\hat\Pi_{11}$:
\begin{equation}\label{eq:pr_single}
 \hat\Pi_{11} \ket{z}_{a_2} \ket{\phi}\ket{z}_{b_2}=\frac{e^{-|z|^2}\alpha}{\sqrt{2}} (\hat a_1^\dagger\hat b_2^\dagger+\hat a_2^\dagger b_1^\dagger )\ket{\Omega},
\end{equation}
where

\begin{multline}
\hat\Pi_{11}=\ket{10;10}\bra{10;10}+\ket{01;10}\bra{01;10}\\
+\ket{01;01}\bra{01;01}+\ket{10;01}\bra{10;01}.
\end{multline}
This projection has Schmidt rank 2 as, in fact, it is (after the normalization) double-rail-encoded path-entangled Bell state. Therefore, after adding ancillary modes condition \eqref{eq:th1} is no longer fulfilled and there is a possibility that one can reveal Bell-nonclassicality of this state using photon number non-mixing projective measurements. This is in fact the case, e.g. see \cite{firstSinglePhoton, SinglePhotonGPY}, where a CH inequality violation is observed for observables totally unrelated to Schmidt decomposition of the state \eqref{eq:single_full}. Let us, however, consider another strategy based on measuring  the projected state \eqref{eq:pr_single}.  One can build upon its Schmidt decomposition the CH inequality of the form \eqref{fieldCH}, which will be maximally ``violated'' if one performs post-selection of results to events with a single photon per side resulting in CH expression value $(\sqrt{2}-1)/2$ for any $z$. Note that if the projected state after normalization would be the full state, then one gets a real violation as one does not have to use post-selection. Measurement which leads to such result could be realized using as observables combinations of Stokes operators projected to single-photon subspace as discussed in example \eqref{eq:bsv_ex}.  However, applying the corresponding CH operator to the full state \eqref{eq:single_full} without using post-selection will result in a lack of violation. This depicts that while one can reveal non-classicality (entanglement) of the state by post-selection to a specific number of particles shared between parties, the same criterion does not have to lead to a proper direct violation of local realism by the actual physical state (post-selection loophole). This additionally shows that generalizations of Gisin's theorem for fixed particle numbers were not sufficient to show that any entangled pure state does not admit LHV model.
\subsection{Violation from projections into specific particle number} 
One could also ask for which states the  CH inequality constructed for some \textit{state projected  into a specific number of particles} will be always  violated by the \textit{total state}.  To answer this question let us note that the lack of violation of the CH inequality for the total state described above comes from modification of local probabilities of measuring an outcome specified in the CH inequality.  More precisely, one can partition the CH operator constructed on the projected state $\ket{\psi_{n,m}}$ into non-local (bipartite)  and two local terms: $\hat{CH}=\hat{CH}_{\textrm{nl}}-\hat{CH}^a_{\textrm{loc}}-\hat{CH}^b_{\textrm{loc}}$ (see Eq. \eqref{fieldCH} and \eqref{probCHop}),  the expectation values of which are modified in different way when the full state is considered. Let us consider projectors  $\hat \Pi_{n'(m')}^{a(b)}$, constructed in analogy to \eqref{eq:sub_p_loc}, which project the state onto given number of particles $n' (m')$ within  groups of modes $\{a_i\}$ ($\{b_l\}$). Additionally let us introduce the notation:
\begin{align}
        \hat\Pi_{n,m'\neq m}&:=\hat \Pi^a_{n}\sum_{m'\neq m}  \hat\Pi^b_{m'},\label{eq:pr_n_m_neq_m}\\
        \hat\Pi_{n'\neq n,m}&:=\hat \Pi^b_{m}\sum_{n'\neq n}  \hat\Pi^a_{n'}.\label{eq:pr_n_neq_n_m}
\end{align}
Note that  $\hat\Pi_{n,m'\neq m} $ stands simply for a projector onto a subspace of states with number of particles different than $m$ in the set of modes $\{b_l\}$ and $n$  particles in the set of modes $\{a_i\}$ and analogously for $\hat\Pi_{n'\neq n,m} $. What is more,  $\sum_n \hat \Pi^a_n$ and $\sum_m \hat \Pi^b_m$ are in fact identity operators on the entire  Fock space. One can easily see that the non-local and local  terms of the CH operator have the following properties:
\begin{align}
    \hat{CH}_{\textrm{nl}}&=\hat \Pi_{nm}\hat{CH}_{\textrm{nl}}\hat \Pi_{nm},\\
    \hat{CH}^{a(b)}_{\textrm{loc}}&=\hat\Pi_{n(m)}^{a(b)}\hat{CH}^{a(b)}_{\textrm{loc}}\hat\Pi_{n(m)}^{a(b)}.\label{eq:ch_loc_eq}
\end{align}
From the first line and \eqref{eq:def_psi} it follows that:
\begin{equation}
\label{CHnl}
    \braket{\hat{CH}_{\textrm{nl}}}_\psi=\braket{\hat\Pi_{nm}}_\psi\braket{\hat{CH}_{\textrm{nl}}}_{\psi_{n,m}}.
\end{equation}
Let us now focus on the expectation value of $ \hat{CH}^{a}_{\textrm{loc}}$ as the considerations for $ \hat{CH}^{b}_{\textrm{loc}}$ follow analogously:
\begin{multline}
\label{CHloc}
  \braket{\hat{CH^a_{\textrm{loc}}}}_{\psi}=\left\langle\hat\Pi_{n}^{a}\sum_{m''} \hat \Pi^b_{m''}\hat{CH}^{a}_{\textrm{loc}}\hat\Pi_{n}^{a}\sum_{m'} \hat \Pi^b_{m'}\right\rangle_{\psi} \\
  =\braket{\hat\Pi_{nm}\hat{CH^a_{\textrm{loc}}}\hat\Pi_{nm}}_{\psi}+\braket{\hat\Pi_{n,m''\neq m}\hat{CH^a_{\textrm{loc}}}\hat\Pi_{n,m'\neq m}}_{\psi}\\
  +\braket{\hat\Pi_{nm}\hat{CH^a_{\textrm{loc}}}\hat\Pi_{n,m'\neq m}}_{\psi}+\braket{\hat\Pi_{n,m''\neq m}\hat{CH^a_{\textrm{loc}}}\hat\Pi_{nm}}_{\psi}\\
  =\braket{\hat\Pi_{nm}}_\psi\braket{\hat{CH}^a_{\textrm{loc}}}_{\psi_{n,m}}+\braket{\hat\Pi_{n,m'\neq m}}_\psi\braket{\hat{CH^a_{\textrm{loc}}}}_{\psi_{n,m'\neq m}},
\end{multline}
where $\braket{...}_{\psi_{n,m'\neq m}}$ stands for expectation value calculated on the state constructed as $\ket{\psi_{n,m}}$ but using projector \eqref{eq:pr_n_m_neq_m}. Here in the first line we have used \eqref{eq:ch_loc_eq} and inserted the identity operator. In the second line we used the fact that $\hat\Pi^a_n\hat\Pi^b_m=\hat\Pi_{nm}$ and definitions \eqref{eq:pr_n_m_neq_m}, \eqref{eq:pr_n_neq_n_m}. In the last equality we have used the fact that the terms in the third line are zero as projectors around $ \hat{CH}^{a}_{\textrm{loc}}$ are orthogonal to each other. One can notice that, while in both expectation values of  $\hat{CH}_{\textrm{nl}}$ and  $ \hat{CH}^{a(b)}_{\textrm{loc}}$  there appear terms corresponding  to rescaled expectation value  of the full CH operator evaluated on the state $\ket{\psi_{n,m}}$,  in the local averages $\braket{\hat{CH}^{a(b)}_{\textrm{loc}}}_{\psi}$ there also appear additional terms which can result in disappearance of the violation. 
To see this let us directly express the average value of the CH operator in terms of its average value on the projected state using formulas \eqref{CHnl} and \eqref{CHloc}:
\begin{align}  
\begin{split} 
&\braket{\hat{CH}}_\psi=\braket{\hat{CH}_{\textrm{nl}}}_\psi- \braket{\hat{CH^a_{\textrm{loc}}}}_{\psi}- \braket{\hat{CH^b_{\textrm{loc}}}}_{\psi}\\
&=\braket{\hat\Pi_{nm}}_\psi\braket{\hat{CH}_{\textrm{nl}}}_{\psi_{n,m}}\\
&-\braket{\hat\Pi_{nm}}_\psi\braket{\hat{CH}^a_{\textrm{loc}}}_{\psi_{n,m}}-\braket{\hat\Pi_{n,m'\neq m}}_\psi\braket{\hat{CH^a_{\textrm{loc}}}}_{\psi_{n,m'\neq m}}\\
&-\braket{\hat\Pi_{nm}}_\psi\braket{\hat{CH}^b_{\textrm{loc}}}_{\psi_{n,m}}-\braket{\hat\Pi_{n'\neq n,m}}_\psi\braket{\hat{CH^b_{\textrm{loc}}}}_{\psi_{n'\neq n,m}}.
\end{split}
\end{align}
The three terms containing $\braket{\hat\Pi_{nm}}_\psi$ sum up to the rescaled average value of the full CH operator evaluated on the projected state,  namely to $\braket{\hat\Pi_{nm}}_\psi\braket{\hat{CH}}_{\psi_{n,m}}$, therefore we finally obtain: 
\begin{align}  
\begin{split} 
&\braket{\hat{CH}}_\psi=\braket{\hat\Pi_{nm}}_\psi\braket{\hat{CH}}_{\psi_{n,m}}\\
&-\braket{\hat\Pi_{n,m'\neq m}}_\psi\braket{\hat{CH^a_{\textrm{loc}}}}_{\psi_{n,m'\neq m}}\\
&-\braket{\hat\Pi_{n'\neq n,m}}_\psi\braket{\hat{CH^b_{\textrm{loc}}}}_{\psi_{n'\neq n,m}}.
\end{split}
\end{align}
Note that the last two terms are always non-positive and result in lowering the average value of the CH operator, which can lead to vanishing of a violation of the inequality. Therefore, the necessary and sufficient condition for violation of the considered type of CH inequality by the full state is:
\begin{multline}
    \exists_{\Pi_{nm}}:\,\,r_{\psi_{n,m}}
>1\, \wedge\, \braket{\hat\Pi_{n,m'\neq m}}_\psi\braket{\hat{CH^a_{\textrm{loc}}}}_{\psi_{n,m'\neq m}}\\ +\braket{\hat\Pi_{n'\neq n,m}}_\psi\braket{\hat{CH^b_{\textrm{loc}}}}_{\psi_{n\neq n',m}}<\braket{\hat\Pi_{nm}}_\psi\braket{\hat{CH}}_{\psi_{n,m}},
\end{multline}
where $r_{\psi_{n,m}}$ stands for the Schmidt rank of the state $\ket{\psi_{n,m}}$.  One can also find a simpler \textit{sufficient} condition:
\begin{equation}
    \exists_{\Pi_{nm}}:\,\,r_{\psi_{n,m}}
>1\, \wedge\, \braket{\hat\Pi_{n'\neq n,m}}_\psi=\braket{\hat\Pi_{n,m'\neq m}}_\psi=0,
\end{equation}
which implies that the state $\ket{\psi}$ has the following easy to identify Schmidt form:
\begin{align}
    \ket{\psi}=&\sum_t\alpha_t F^n(a_i^\dagger,t)G^m(b_l^\dagger,t)\ket{\Omega}\\
    &+\sum_{t'}\alpha'_{t'} F'(a_i^\dagger,t')G'(b_l^\dagger,t')\ket{\Omega},
\end{align}
where $F'(a_i^\dagger,t'),\,G'(b_l^\dagger,t')$ are polynomials of the corresponding creation operators such that:
\begin{equation}
    \braket{\Omega|F'(a_i^\dagger,t')^\dagger G'(b_l^\dagger,t')^\dagger\hat\Pi^{a(b)}_{n(m)} F'(a_i^\dagger,t')G'(b_l^\dagger,t')|\Omega}=0.
\end{equation}
Such a form, in fact, occurred in the examples of BGHZ \eqref{BGHZ0}  and BSV \eqref{eq:bsv} states (for those states $\forall_{t'}\, \alpha'_{t'}=0$). 
\section{Concluding remarks}
In summary, in this work we have proved that any pure mode-entangled state of quantum fields leads to a violation of some Bell inequality and, therefore, we provide a generalization of Gisin's theorem to quantum fields.
This result also generalizes to non-separable Hilbert spaces used in thermodynamic limits of quantum field theory.  What is more, we provided sufficient condition for the experimental need of introduction of ancillary resources out of a passive linear optics paradigm in order to measure Bell-inequality violation. Finally, we have discussed several examples among which of particular importance are two mode squeezed vacuum and beam-splitted single-photon as entanglement of those states is distributed over states with different particle numbers (any of the cuts for specific particle number is not entangled), thus those are not covered by previous iterations of generalizations of Gisin's theorem.

Let us make a remark about the fundamental consequences of these results. While the result presented provides the most general form of Gisin's theorem in the currently established theoretical quantum model of nature (quantum field theory), this does not give the final answer to the question if any entanglement  remains in contradiction with local realism. This is because  we still  do not have unified quantum theory involving gravity, and the consequences of such construction do not have to lead to exactly the same predictions in general. However, this result shows that quantum field theory predicts that any quantum measurement understood in the sense of standard quantum measurement theory \cite{Busch91, Zurek.82, Zurek.03} cannot be associated with the appearance of some local hidden variable (local definite record of the outcome). This is because unitary evolution of the full system will lead to entanglement of the measured subsystem with the environment (measurement apparatus) trading for it inside coherences of the subsystem.
However, as long as the entanglement is present, the local hidden variable cannot appear.  Still, one cannot be sure if more general theory would follow such predictions. 

%In order to local hidden variable to appear as a part of measurement the global evolution would need to be non-unitary what leads to ideas like objective collapse theories. Still,

\begin{acknowledgements}
The work is part of ‘International Centre for
Theory of Quantum Technologies’ project (contract no. 2018/MAB/5), which is carried out
within the International Research Agendas Programme (IRAP) of the Foundation for Polish
Science (FNP) co-financed by the European Union from the funds of the Smart Growth
Operational Programme, axis IV: Increasing the research potential (Measure 4.3). The authors  acknowledge discussions with  Marek \.Zukowski, Bianka Wo\l{}oncewicz, Tamoghna Das, Shilpa Samaddar and Antonio Mandarino.
\end{acknowledgements}

\appendix
\section{Schmidt decomposition in non-separable Hilbert space}\label{app:A}
One can build the Schmidt decomposition of a state in a non-separable Hilbert space $\mathcal{H}'$  directly from the non-countable extended occupation number basis  $\ket{\vec n^j; \vec m^{j'}}$ without restrictions $\sum^{\overline{\{a_i\}}}_{i=1} n^j_i<\infty, \,\sum^{\overline{\{b_{l}\}}}_{l=1}m^{j'}_{l}<\infty$. Let us recall that any state to which one can properly assign  probabilities of outcomes has a non-zero distribution of probability only in the separable subspace $\mathcal K_\psi$ of $\mathcal{H}'$ and is equivalent to a state  $\ket{\psi}$ from this subspace. One can build a countable basis $\ket{k,k'}$ in the subspace $\mathcal K_\psi$ using basis vectors $\ket{\vec n^j; \vec m^{j'}}$ of $\mathcal{H}'$ as:
\begin{equation}
    \ket{k,k'}=\int dj \beta_{k}(j) \int dj'\gamma_{k'}(j') \ket{\vec n^j; \vec m^{j'}},
\end{equation}
where now $j,j'$ are continuous indices, and $\beta_{k}(j),\gamma_{k'}(j')$ are amplitudes given by generalized functions. Note that these integrals could be, in some circumstances, reduced to sums over  countable indices when amplitudes are given by countable sums of Dirac deltas. 
Then again one builds two abstract separable Hilbert spaces $\mathcal{H}_{A(B)}$ with one basis vector $\tilde\beta_k\, (\tilde\gamma_k')$  for each generalized function $\beta_{k}(j),\gamma_{k'}(j')$. Upon this, one builds isomorphism $\mathcal{K}_\psi\rightarrow\mathcal{H}_{A}\otimes \mathcal{H}_{B}$ as $\ket{k,k'}\rightarrow \tilde\beta_k\otimes\tilde\gamma_k'$. Then one can again perform decomposition of  $\ket{\psi}$ in $\mathcal{H}_{A}\otimes \mathcal{H}_{B}$ and transform it back to  $\mathcal{K}_\psi$ reaching the decomposition of the form:
\begin{widetext}
\begin{equation}
\ket{\psi}=\sum_{t}\alpha_t\sum_{k}\xi_k^t\sum_{k'}\zeta_{k'}^t\ket{k,k'}\\=\sum_{t}\alpha_t\left[\int dj\left(\sum_{k}\xi_k^t \beta_{k}\right)\int dj'\left(\sum_{k'}\zeta_{k'}^t\gamma_{k'}(j')\right)\ket{\vec n^j; \vec m^{j'}}\right].
\end{equation}
\end{widetext}

\begin{thebibliography}{44}%
    \makeatletter
    \providecommand \@ifxundefined [1]{%
     \@ifx{#1\undefined}
    }%
    \providecommand \@ifnum [1]{%
     \ifnum #1\expandafter \@firstoftwo
     \else \expandafter \@secondoftwo
     \fi
    }%
    \providecommand \@ifx [1]{%
     \ifx #1\expandafter \@firstoftwo
     \else \expandafter \@secondoftwo
     \fi
    }%
    \providecommand \natexlab [1]{#1}%
    \providecommand \enquote  [1]{``#1''}%
    \providecommand \bibnamefont  [1]{#1}%
    \providecommand \bibfnamefont [1]{#1}%
    \providecommand \citenamefont [1]{#1}%
    \providecommand \href@noop [0]{\@secondoftwo}%
    \providecommand \href [0]{\begingroup \@sanitize@url \@href}%
    \providecommand \@href[1]{\@@startlink{#1}\@@href}%
    \providecommand \@@href[1]{\endgroup#1\@@endlink}%
    \providecommand \@sanitize@url [0]{\catcode `\\12\catcode `\$12\catcode
      `\&12\catcode `\#12\catcode `\^12\catcode `\_12\catcode `\%12\relax}%
    \providecommand \@@startlink[1]{}%
    \providecommand \@@endlink[0]{}%
    \providecommand \url  [0]{\begingroup\@sanitize@url \@url }%
    \providecommand \@url [1]{\endgroup\@href {#1}{\urlprefix }}%
    \providecommand \urlprefix  [0]{URL }%
    \providecommand \Eprint [0]{\href }%
    \providecommand \doibase [0]{https://doi.org/}%
    \providecommand \selectlanguage [0]{\@gobble}%
    \providecommand \bibinfo  [0]{\@secondoftwo}%
    \providecommand \bibfield  [0]{\@secondoftwo}%
    \providecommand \translation [1]{[#1]}%
    \providecommand \BibitemOpen [0]{}%
    \providecommand \bibitemStop [0]{}%
    \providecommand \bibitemNoStop [0]{.\EOS\space}%
    \providecommand \EOS [0]{\spacefactor3000\relax}%
    \providecommand \BibitemShut  [1]{\csname bibitem#1\endcsname}%
    \let\auto@bib@innerbib\@empty
    %</preamble>
    \bibitem [{\citenamefont {Einstein}\ \emph {et~al.}(1935)\citenamefont
      {Einstein}, \citenamefont {Podolsky},\ and\ \citenamefont {Rosen}}]{EPR}%
      \BibitemOpen
      \bibfield  {author} {\bibinfo {author} {\bibfnamefont {A.}~\bibnamefont
      {Einstein}}, \bibinfo {author} {\bibfnamefont {B.}~\bibnamefont {Podolsky}},\
      and\ \bibinfo {author} {\bibfnamefont {N.}~\bibnamefont {Rosen}},\ }\bibfield
       {title} {\bibinfo {title} {Can quantum-mechanical description of physical
      reality be considered complete?},\ }\href
      {https://doi.org/10.1103/PhysRev.47.777} {\bibfield  {journal} {\bibinfo
      {journal} {Phys. Rev.}\ }\textbf {\bibinfo {volume} {47}},\ \bibinfo {pages}
      {777} (\bibinfo {year} {1935})}\BibitemShut {NoStop}%
    \bibitem [{\citenamefont {Aspect}(2002)}]{Aspect2002}%
      \BibitemOpen
      \bibfield  {author} {\bibinfo {author} {\bibfnamefont {A.}~\bibnamefont
      {Aspect}},\ }\bibinfo {title} {Bell's theorem: The naive view of an
      experimentalist},\ in\ \href {https://doi.org/10.1007/978-3-662-05032-3_9}
      {\emph {\bibinfo {booktitle} {Quantum [Un]speakables: From Bell to Quantum
      Information}}}\ (\bibinfo  {publisher} {Springer Berlin Heidelberg},\
      \bibinfo {address} {Berlin, Heidelberg},\ \bibinfo {year} {2002})\ pp.\
      \bibinfo {pages} {119--153}\BibitemShut {NoStop}%
    \bibitem [{\citenamefont {Brukner}\ and\ \citenamefont
      {{\.{Z}}ukowski}(2012)}]{Brukner2012}%
      \BibitemOpen
      \bibfield  {author} {\bibinfo {author} {\bibfnamefont {{\v{C}}.}~\bibnamefont
      {Brukner}}\ and\ \bibinfo {author} {\bibfnamefont {M.}~\bibnamefont
      {{\.{Z}}ukowski}},\ }\bibinfo {title} {Bell's inequalities --- foundations
      and quantum communication},\ in\ \href
      {https://doi.org/10.1007/978-3-540-92910-9_42} {\emph {\bibinfo {booktitle}
      {Handbook of Natural Computing}}},\ \bibinfo {editor} {edited by\ \bibinfo
      {editor} {\bibfnamefont {G.}~\bibnamefont {Rozenberg}}, \bibinfo {editor}
      {\bibfnamefont {T.}~\bibnamefont {B{\"a}ck}},\ and\ \bibinfo {editor}
      {\bibfnamefont {J.~N.}\ \bibnamefont {Kok}}}\ (\bibinfo  {publisher}
      {Springer Berlin Heidelberg},\ \bibinfo {address} {Berlin, Heidelberg},\
      \bibinfo {year} {2012})\ pp.\ \bibinfo {pages} {1413--1450}\BibitemShut
      {NoStop}%
    \bibitem [{\citenamefont {Pan}\ \emph {et~al.}(2012)\citenamefont {Pan},
      \citenamefont {Chen}, \citenamefont {Lu}, \citenamefont {Weinfurter},
      \citenamefont {Zeilinger},\ and\ \citenamefont {\ifmmode~\dot{Z}\else
      \.{Z}\fi{}ukowski}}]{Pan2012}%
      \BibitemOpen
      \bibfield  {author} {\bibinfo {author} {\bibfnamefont {J.-W.}\ \bibnamefont
      {Pan}}, \bibinfo {author} {\bibfnamefont {Z.-B.}\ \bibnamefont {Chen}},
      \bibinfo {author} {\bibfnamefont {C.-Y.}\ \bibnamefont {Lu}}, \bibinfo
      {author} {\bibfnamefont {H.}~\bibnamefont {Weinfurter}}, \bibinfo {author}
      {\bibfnamefont {A.}~\bibnamefont {Zeilinger}},\ and\ \bibinfo {author}
      {\bibfnamefont {M.}~\bibnamefont {\ifmmode~\dot{Z}\else \.{Z}\fi{}ukowski}},\
      }\bibfield  {title} {\bibinfo {title} {Multiphoton entanglement and
      interferometry},\ }\href {https://doi.org/10.1103/RevModPhys.84.777}
      {\bibfield  {journal} {\bibinfo  {journal} {Rev. Mod. Phys.}\ }\textbf
      {\bibinfo {volume} {84}},\ \bibinfo {pages} {777} (\bibinfo {year}
      {2012})}\BibitemShut {NoStop}%
    \bibitem [{\citenamefont {Werner}\ and\ \citenamefont
      {Wolf}(2001)}]{Werner2001}%
      \BibitemOpen
      \bibfield  {author} {\bibinfo {author} {\bibfnamefont {R.~F.}\ \bibnamefont
      {Werner}}\ and\ \bibinfo {author} {\bibfnamefont {M.~M.}\ \bibnamefont
      {Wolf}},\ }\bibfield  {title} {\bibinfo {title} {Bell inequalities and
      entanglement},\ }\href {https://doi.org/10.26421/QIC1.3-1} {\bibfield
      {journal} {\bibinfo  {journal} {QIC}\ }\textbf {\bibinfo {volume} {1}},\
      \bibinfo {pages} {1} (\bibinfo {year} {2001})}\BibitemShut {NoStop}%
    \bibitem [{\citenamefont {\ifmmode~\dot{Z}\else \.{Z}\fi{}ukowski}\ and\
      \citenamefont {Brukner}(2002)}]{Zukowski2002}%
      \BibitemOpen
      \bibfield  {author} {\bibinfo {author} {\bibfnamefont {M.}~\bibnamefont
      {\ifmmode~\dot{Z}\else \.{Z}\fi{}ukowski}}\ and\ \bibinfo {author}
      {\bibfnamefont {{\v{C}}.}~\bibnamefont {Brukner}},\ }\bibfield  {title}
      {\bibinfo {title} {{Bell's Theorem for General N-Qubit States}},\ }\href
      {https://doi.org/10.1103/PhysRevLett.88.210401} {\bibfield  {journal}
      {\bibinfo  {journal} {Phys. Rev. Lett.}\ }\textbf {\bibinfo {volume} {88}},\
      \bibinfo {pages} {210401} (\bibinfo {year} {2002})}\BibitemShut {NoStop}%
    \bibitem [{\citenamefont {Brunner}\ \emph {et~al.}(2014)\citenamefont
      {Brunner}, \citenamefont {Cavalcanti}, \citenamefont {Pironio}, \citenamefont
      {Scarani},\ and\ \citenamefont {Wehner}}]{BrunnerRev}%
      \BibitemOpen
      \bibfield  {author} {\bibinfo {author} {\bibfnamefont {N.}~\bibnamefont
      {Brunner}}, \bibinfo {author} {\bibfnamefont {D.}~\bibnamefont {Cavalcanti}},
      \bibinfo {author} {\bibfnamefont {S.}~\bibnamefont {Pironio}}, \bibinfo
      {author} {\bibfnamefont {V.}~\bibnamefont {Scarani}},\ and\ \bibinfo {author}
      {\bibfnamefont {S.}~\bibnamefont {Wehner}},\ }\bibfield  {title} {\bibinfo
      {title} {Bell nonlocality},\ }\href
      {https://doi.org/10.1103/RevModPhys.86.419} {\bibfield  {journal} {\bibinfo
      {journal} {Rev. Mod. Phys.}\ }\textbf {\bibinfo {volume} {86}},\ \bibinfo
      {pages} {419} (\bibinfo {year} {2014})}\BibitemShut {NoStop}%
    \bibitem [{\citenamefont {Bell}(1964)}]{BELL}%
      \BibitemOpen
      \bibfield  {author} {\bibinfo {author} {\bibfnamefont {J.~S.}\ \bibnamefont
      {Bell}},\ }\bibfield  {title} {\bibinfo {title} {On the {Einstein} {Podolsky}
      {Rosen} paradox},\ }\href
      {https://doi.org/10.1103/PhysicsPhysiqueFizika.1.195} {\bibfield  {journal}
      {\bibinfo  {journal} {Physics Physique Fizika}\ }\textbf {\bibinfo {volume}
      {1}},\ \bibinfo {pages} {195} (\bibinfo {year} {1964})}\BibitemShut {NoStop}%
    \bibitem [{\citenamefont {Aspect}\ \emph {et~al.}(1982)\citenamefont {Aspect},
      \citenamefont {Dalibard},\ and\ \citenamefont {Roger}}]{Aspect1982}%
      \BibitemOpen
      \bibfield  {author} {\bibinfo {author} {\bibfnamefont {A.}~\bibnamefont
      {Aspect}}, \bibinfo {author} {\bibfnamefont {J.}~\bibnamefont {Dalibard}},\
      and\ \bibinfo {author} {\bibfnamefont {G.}~\bibnamefont {Roger}},\ }\bibfield
       {title} {\bibinfo {title} {Experimental test of {Bell's} inequalities using
      time-varying analyzers},\ }\href
      {https://doi.org/10.1103/PhysRevLett.49.1804} {\bibfield  {journal} {\bibinfo
       {journal} {Phys. Rev. Lett.}\ }\textbf {\bibinfo {volume} {49}},\ \bibinfo
      {pages} {1804} (\bibinfo {year} {1982})}\BibitemShut {NoStop}%
    \bibitem [{\citenamefont {Hensen}\ \emph {et~al.}(2015)\citenamefont {Hensen},
      \citenamefont {Bernien}, \citenamefont {Dr{\'e}au}, \citenamefont {Reiserer},
      \citenamefont {Kalb}, \citenamefont {Blok}, \citenamefont {Ruitenberg},
      \citenamefont {Vermeulen}, \citenamefont {Schouten}, \citenamefont
      {Abell{\'a}n}, \citenamefont {Amaya}, \citenamefont {Pruneri}, \citenamefont
      {Mitchell}, \citenamefont {Markham}, \citenamefont {Twitchen}, \citenamefont
      {Elkouss}, \citenamefont {Wehner}, \citenamefont {Taminiau},\ and\
      \citenamefont {Hanson}}]{Hensen2015}%
      \BibitemOpen
      \bibfield  {author} {\bibinfo {author} {\bibfnamefont {B.}~\bibnamefont
      {Hensen}}, \bibinfo {author} {\bibfnamefont {H.}~\bibnamefont {Bernien}},
      \bibinfo {author} {\bibfnamefont {A.~E.}\ \bibnamefont {Dr{\'e}au}}, \bibinfo
      {author} {\bibfnamefont {A.}~\bibnamefont {Reiserer}}, \bibinfo {author}
      {\bibfnamefont {N.}~\bibnamefont {Kalb}}, \bibinfo {author} {\bibfnamefont
      {M.~S.}\ \bibnamefont {Blok}}, \bibinfo {author} {\bibfnamefont
      {J.}~\bibnamefont {Ruitenberg}}, \bibinfo {author} {\bibfnamefont {R.~F.~L.}\
      \bibnamefont {Vermeulen}}, \bibinfo {author} {\bibfnamefont {R.~N.}\
      \bibnamefont {Schouten}}, \bibinfo {author} {\bibfnamefont {C.}~\bibnamefont
      {Abell{\'a}n}}, \bibinfo {author} {\bibfnamefont {W.}~\bibnamefont {Amaya}},
      \bibinfo {author} {\bibfnamefont {V.}~\bibnamefont {Pruneri}}, \bibinfo
      {author} {\bibfnamefont {M.~W.}\ \bibnamefont {Mitchell}}, \bibinfo {author}
      {\bibfnamefont {M.}~\bibnamefont {Markham}}, \bibinfo {author} {\bibfnamefont
      {D.~J.}\ \bibnamefont {Twitchen}}, \bibinfo {author} {\bibfnamefont
      {D.}~\bibnamefont {Elkouss}}, \bibinfo {author} {\bibfnamefont
      {S.}~\bibnamefont {Wehner}}, \bibinfo {author} {\bibfnamefont {T.~H.}\
      \bibnamefont {Taminiau}},\ and\ \bibinfo {author} {\bibfnamefont
      {R.}~\bibnamefont {Hanson}},\ }\bibfield  {title} {\bibinfo {title}
      {Loophole-free {Bell} inequality violation using electron spins separated by
      1.3 kilometres},\ }\href {https://doi.org/10.1038/nature15759} {\bibfield
      {journal} {\bibinfo  {journal} {Nature}\ }\textbf {\bibinfo {volume} {526}},\
      \bibinfo {pages} {682} (\bibinfo {year} {2015})}\BibitemShut {NoStop}%
    \bibitem [{\citenamefont {Gisin}(1991)}]{Gisin_PLA}%
      \BibitemOpen
      \bibfield  {author} {\bibinfo {author} {\bibfnamefont {N.}~\bibnamefont
      {Gisin}},\ }\bibfield  {title} {\bibinfo {title} {Bell's inequality holds for
      all non-product states},\ }\href
      {https://doi.org/https://doi.org/10.1016/0375-9601(91)90805-I} {\bibfield
      {journal} {\bibinfo  {journal} {Physics Letters A}\ }\textbf {\bibinfo
      {volume} {154}},\ \bibinfo {pages} {201} (\bibinfo {year}
      {1991})}\BibitemShut {NoStop}%
    \bibitem [{\citenamefont {Popescu}\ and\ \citenamefont
      {Rohrlich}(1992)}]{POPESCU}%
      \BibitemOpen
      \bibfield  {author} {\bibinfo {author} {\bibfnamefont {S.}~\bibnamefont
      {Popescu}}\ and\ \bibinfo {author} {\bibfnamefont {D.}~\bibnamefont
      {Rohrlich}},\ }\bibfield  {title} {\bibinfo {title} {Generic quantum
      nonlocality},\ }\href
      {https://doi.org/https://doi.org/10.1016/0375-9601(92)90711-T} {\bibfield
      {journal} {\bibinfo  {journal} {Physics Letters A}\ }\textbf {\bibinfo
      {volume} {166}},\ \bibinfo {pages} {293} (\bibinfo {year}
      {1992})}\BibitemShut {NoStop}%
    \bibitem [{\citenamefont {Gachechiladze}\ and\ \citenamefont
      {Gühne}(2017)}]{GACHECHILADZE}%
      \BibitemOpen
      \bibfield  {author} {\bibinfo {author} {\bibfnamefont {M.}~\bibnamefont
      {Gachechiladze}}\ and\ \bibinfo {author} {\bibfnamefont {O.}~\bibnamefont
      {Gühne}},\ }\bibfield  {title} {\bibinfo {title} {Completing the proof of
      “generic quantum nonlocality”},\ }\href
      {https://doi.org/https://doi.org/10.1016/j.physleta.2016.10.001} {\bibfield
      {journal} {\bibinfo  {journal} {Physics Letters A}\ }\textbf {\bibinfo
      {volume} {381}},\ \bibinfo {pages} {1281} (\bibinfo {year}
      {2017})}\BibitemShut {NoStop}%
    \bibitem [{\citenamefont {Li}\ and\ \citenamefont {Fei}(2010)}]{Gisin_inf2}%
      \BibitemOpen
      \bibfield  {author} {\bibinfo {author} {\bibfnamefont {M.}~\bibnamefont
      {Li}}\ and\ \bibinfo {author} {\bibfnamefont {S.-M.}\ \bibnamefont {Fei}},\
      }\bibfield  {title} {\bibinfo {title} {Gisin's theorem for arbitrary
      dimensional multipartite states},\ }\href
      {https://doi.org/10.1103/PhysRevLett.104.240502} {\bibfield  {journal}
      {\bibinfo  {journal} {Phys. Rev. Lett.}\ }\textbf {\bibinfo {volume} {104}},\
      \bibinfo {pages} {240502} (\bibinfo {year} {2010})}\BibitemShut {NoStop}%
    \bibitem [{\citenamefont {Yu}\ \emph {et~al.}(2012)\citenamefont {Yu},
      \citenamefont {Chen}, \citenamefont {Zhang}, \citenamefont {Lai},\ and\
      \citenamefont {Oh}}]{Gisin_inf}%
      \BibitemOpen
      \bibfield  {author} {\bibinfo {author} {\bibfnamefont {S.}~\bibnamefont
      {Yu}}, \bibinfo {author} {\bibfnamefont {Q.}~\bibnamefont {Chen}}, \bibinfo
      {author} {\bibfnamefont {C.}~\bibnamefont {Zhang}}, \bibinfo {author}
      {\bibfnamefont {C.~H.}\ \bibnamefont {Lai}},\ and\ \bibinfo {author}
      {\bibfnamefont {C.~H.}\ \bibnamefont {Oh}},\ }\bibfield  {title} {\bibinfo
      {title} {All entangled pure states violate a single {Bell's} inequality},\
      }\href {https://doi.org/10.1103/PhysRevLett.109.120402} {\bibfield  {journal}
      {\bibinfo  {journal} {Phys. Rev. Lett.}\ }\textbf {\bibinfo {volume} {109}},\
      \bibinfo {pages} {120402} (\bibinfo {year} {2012})}\BibitemShut {NoStop}%
    \bibitem [{\citenamefont {Chekhova}\ \emph {et~al.}(2015)\citenamefont
      {Chekhova}, \citenamefont {Leuchs},\ and\ \citenamefont {Żukowski}}]{Masza}%
      \BibitemOpen
      \bibfield  {author} {\bibinfo {author} {\bibfnamefont {M.}~\bibnamefont
      {Chekhova}}, \bibinfo {author} {\bibfnamefont {G.}~\bibnamefont {Leuchs}},\
      and\ \bibinfo {author} {\bibfnamefont {M.}~\bibnamefont {Żukowski}},\
      }\bibfield  {title} {\bibinfo {title} {Bright squeezed vacuum: Entanglement
      of macroscopic light beams},\ }\href
      {https://doi.org/http://doi.org/10.1016/j.optcom.2014.07.050} {\bibfield
      {journal} {\bibinfo  {journal} {Optics Communications}\ }\textbf {\bibinfo
      {volume} {337}},\ \bibinfo {pages} {27} (\bibinfo {year} {2015})}\BibitemShut
      {NoStop}%
    \bibitem [{\citenamefont {Hillery}\ and\ \citenamefont
      {Zubairy}(2006)}]{two_mode_ent}%
      \BibitemOpen
      \bibfield  {author} {\bibinfo {author} {\bibfnamefont {M.}~\bibnamefont
      {Hillery}}\ and\ \bibinfo {author} {\bibfnamefont {M.~S.}\ \bibnamefont
      {Zubairy}},\ }\bibfield  {title} {\bibinfo {title} {Entanglement conditions
      for two-mode states},\ }\href {https://doi.org/10.1103/PhysRevLett.96.050503}
      {\bibfield  {journal} {\bibinfo  {journal} {Phys. Rev. Lett.}\ }\textbf
      {\bibinfo {volume} {96}},\ \bibinfo {pages} {050503} (\bibinfo {year}
      {2006})}\BibitemShut {NoStop}%
    \bibitem [{\citenamefont {Hardy}(1994)}]{Hardy94}%
      \BibitemOpen
      \bibfield  {author} {\bibinfo {author} {\bibfnamefont {L.}~\bibnamefont
      {Hardy}},\ }\bibfield  {title} {\bibinfo {title} {Nonlocality of a single
      photon revisited},\ }\href {https://doi.org/10.1103/PhysRevLett.73.2279}
      {\bibfield  {journal} {\bibinfo  {journal} {Phys. Rev. Lett.}\ }\textbf
      {\bibinfo {volume} {73}},\ \bibinfo {pages} {2279} (\bibinfo {year}
      {1994})}\BibitemShut {NoStop}%
    \bibitem [{\citenamefont {van Enk}(2005)}]{Enk05}%
      \BibitemOpen
      \bibfield  {author} {\bibinfo {author} {\bibfnamefont {S.~J.}\ \bibnamefont
      {van Enk}},\ }\bibfield  {title} {\bibinfo {title} {Single-particle
      entanglement},\ }\href {https://doi.org/10.1103/PhysRevA.72.064306}
      {\bibfield  {journal} {\bibinfo  {journal} {Phys. Rev. A}\ }\textbf {\bibinfo
      {volume} {72}},\ \bibinfo {pages} {064306} (\bibinfo {year}
      {2005})}\BibitemShut {NoStop}%
    \bibitem [{\citenamefont {Clauser}\ and\ \citenamefont {Horne}(1974)}]{CH}%
      \BibitemOpen
      \bibfield  {author} {\bibinfo {author} {\bibfnamefont {J.~F.}\ \bibnamefont
      {Clauser}}\ and\ \bibinfo {author} {\bibfnamefont {M.~A.}\ \bibnamefont
      {Horne}},\ }\bibfield  {title} {\bibinfo {title} {Experimental consequences
      of objective local theories},\ }\href
      {https://doi.org/10.1103/PhysRevD.10.526} {\bibfield  {journal} {\bibinfo
      {journal} {Phys. Rev. D}\ }\textbf {\bibinfo {volume} {10}},\ \bibinfo
      {pages} {526} (\bibinfo {year} {1974})}\BibitemShut {NoStop}%
    \bibitem [{\citenamefont {Reed}\ and\ \citenamefont {Simon}(1981)}]{reed1981}%
      \BibitemOpen
      \bibfield  {author} {\bibinfo {author} {\bibfnamefont {M.}~\bibnamefont
      {Reed}}\ and\ \bibinfo {author} {\bibfnamefont {B.}~\bibnamefont {Simon}},\
      }\href {https://books.google.pl/books?id=rpFTTjxOYpsC} {\emph {\bibinfo
      {title} {I: Functional Analysis}}},\ Methods of Modern Mathematical Physics\
      (\bibinfo  {publisher} {Elsevier Science},\ \bibinfo {year} {1981})\ \bibinfo
      {note} {p.220}\BibitemShut {NoStop}%
    \bibitem [{\citenamefont {Wiseman}\ and\ \citenamefont
      {Vaccaro}(2003)}]{Wiseman03}%
      \BibitemOpen
      \bibfield  {author} {\bibinfo {author} {\bibfnamefont {H.~M.}\ \bibnamefont
      {Wiseman}}\ and\ \bibinfo {author} {\bibfnamefont {J.~A.}\ \bibnamefont
      {Vaccaro}},\ }\bibfield  {title} {\bibinfo {title} {Entanglement of
      indistinguishable particles shared between two parties},\ }\href
      {https://doi.org/10.1103/PhysRevLett.91.097902} {\bibfield  {journal}
      {\bibinfo  {journal} {Phys. Rev. Lett.}\ }\textbf {\bibinfo {volume} {91}},\
      \bibinfo {pages} {097902} (\bibinfo {year} {2003})}\BibitemShut {NoStop}%
    \bibitem [{\citenamefont {Demkowicz-Dobrzański}\ \emph
      {et~al.}(2015)\citenamefont {Demkowicz-Dobrzański}, \citenamefont
      {Jarzyna},\ and\ \citenamefont {Kołodyński}}]{Demkowicz15}%
      \BibitemOpen
      \bibfield  {author} {\bibinfo {author} {\bibfnamefont {R.}~\bibnamefont
      {Demkowicz-Dobrzański}}, \bibinfo {author} {\bibfnamefont {M.}~\bibnamefont
      {Jarzyna}},\ and\ \bibinfo {author} {\bibfnamefont {J.}~\bibnamefont
      {Kołodyński}},\ }\href
      {https://doi.org/https://doi.org/10.1016/bs.po.2015.02.003} {\emph {\bibinfo
      {title} {Chapter Four - Quantum Limits in Optical Interferometry}}},\ edited
      by\ \bibinfo {editor} {\bibfnamefont {E.}~\bibnamefont {Wolf}},\ \bibinfo
      {series} {Progress in Optics}, Vol.~\bibinfo {volume} {60}\ (\bibinfo
      {publisher} {Elsevier},\ \bibinfo {year} {2015})\ pp.\ \bibinfo {pages} {345
      -- 435}\BibitemShut {NoStop}%
    \bibitem [{\citenamefont {Benatti}\ \emph {et~al.}(2020)\citenamefont
      {Benatti}, \citenamefont {Floreanini}, \citenamefont {Franchini},\ and\
      \citenamefont {Marzolino}}]{BENATTI20201}%
      \BibitemOpen
      \bibfield  {author} {\bibinfo {author} {\bibfnamefont {F.}~\bibnamefont
      {Benatti}}, \bibinfo {author} {\bibfnamefont {R.}~\bibnamefont {Floreanini}},
      \bibinfo {author} {\bibfnamefont {F.}~\bibnamefont {Franchini}},\ and\
      \bibinfo {author} {\bibfnamefont {U.}~\bibnamefont {Marzolino}},\ }\bibfield
      {title} {\bibinfo {title} {Entanglement in indistinguishable particle
      systems},\ }\href
      {https://doi.org/https://doi.org/10.1016/j.physrep.2020.07.003} {\bibfield
      {journal} {\bibinfo  {journal} {Physics Reports}\ }\textbf {\bibinfo {volume}
      {878}},\ \bibinfo {pages} {1} (\bibinfo {year} {2020})}\BibitemShut {NoStop}%
    \bibitem [{\citenamefont {Blasiak}\ and\ \citenamefont
      {Markiewicz}(2019)}]{Blasiak19}%
      \BibitemOpen
      \bibfield  {author} {\bibinfo {author} {\bibfnamefont {P.}~\bibnamefont
      {Blasiak}}\ and\ \bibinfo {author} {\bibfnamefont {M.}~\bibnamefont
      {Markiewicz}},\ }\bibfield  {title} {\bibinfo {title} {Entangling three
      qubits without ever touching},\ }\href
      {https://doi.org/10.1038/s41598-019-55137-3} {\bibfield  {journal} {\bibinfo
      {journal} {Scientific Reports}\ }\textbf {\bibinfo {volume} {9}},\ \bibinfo
      {pages} {20131} (\bibinfo {year} {2019})}\BibitemShut {NoStop}%
    \bibitem [{\citenamefont {Blasiak}\ \emph {et~al.}(2021)\citenamefont
      {Blasiak}, \citenamefont {Borsuk}, \citenamefont {Markiewicz},\ and\
      \citenamefont {Kim}}]{Blasiak21}%
      \BibitemOpen
      \bibfield  {author} {\bibinfo {author} {\bibfnamefont {P.}~\bibnamefont
      {Blasiak}}, \bibinfo {author} {\bibfnamefont {E.}~\bibnamefont {Borsuk}},
      \bibinfo {author} {\bibfnamefont {M.}~\bibnamefont {Markiewicz}},\ and\
      \bibinfo {author} {\bibfnamefont {Y.-S.}\ \bibnamefont {Kim}},\ }\bibfield
      {title} {\bibinfo {title} {Efficient linear-optical generation of a
      multipartite {$W$} state},\ }\href
      {https://doi.org/10.1103/PhysRevA.104.023701} {\bibfield  {journal} {\bibinfo
       {journal} {Phys. Rev. A}\ }\textbf {\bibinfo {volume} {104}},\ \bibinfo
      {pages} {023701} (\bibinfo {year} {2021})}\BibitemShut {NoStop}%
    \bibitem [{\citenamefont {Blasiak}\ \emph {et~al.}(2022)\citenamefont
      {Blasiak}, \citenamefont {Borsuk},\ and\ \citenamefont
      {Markiewicz}}]{Blasiak22}%
      \BibitemOpen
      \bibfield  {author} {\bibinfo {author} {\bibfnamefont {P.}~\bibnamefont
      {Blasiak}}, \bibinfo {author} {\bibfnamefont {E.}~\bibnamefont {Borsuk}},\
      and\ \bibinfo {author} {\bibfnamefont {M.}~\bibnamefont {Markiewicz}},\
      }\bibfield  {title} {\bibinfo {title} {Arbitrary entanglement of three qubits
      via linear optics},\ }\href {https://doi.org/10.1038/s41598-022-22835-4}
      {\bibfield  {journal} {\bibinfo  {journal} {Scientific Reports}\ }\textbf
      {\bibinfo {volume} {12}},\ \bibinfo {pages} {21596} (\bibinfo {year}
      {2022})}\BibitemShut {NoStop}%
    \bibitem [{\citenamefont {Choi}(1975)}]{Choi75}%
      \BibitemOpen
      \bibfield  {author} {\bibinfo {author} {\bibfnamefont {M.-D.}\ \bibnamefont
      {Choi}},\ }\bibfield  {title} {\bibinfo {title} {Completely positive linear
      maps on complex matrices},\ }\href
      {https://doi.org/https://doi.org/10.1016/0024-3795(75)90075-0} {\bibfield
      {journal} {\bibinfo  {journal} {Linear Algebra and its Applications}\
      }\textbf {\bibinfo {volume} {10}},\ \bibinfo {pages} {285} (\bibinfo {year}
      {1975})}\BibitemShut {NoStop}%
    \bibitem [{\citenamefont {Jamiołkowski}(1972)}]{Jamiolkowski72}%
      \BibitemOpen
      \bibfield  {author} {\bibinfo {author} {\bibfnamefont {A.}~\bibnamefont
      {Jamiołkowski}},\ }\bibfield  {title} {\bibinfo {title} {Linear
      transformations which preserve trace and positive semidefiniteness of
      operators},\ }\href
      {https://doi.org/https://doi.org/10.1016/0034-4877(72)90011-0} {\bibfield
      {journal} {\bibinfo  {journal} {Reports on Mathematical Physics}\ }\textbf
      {\bibinfo {volume} {3}},\ \bibinfo {pages} {275} (\bibinfo {year}
      {1972})}\BibitemShut {NoStop}%
    \bibitem [{\citenamefont {Scarani}(2019)}]{CH_eq}%
      \BibitemOpen
      \bibfield  {author} {\bibinfo {author} {\bibfnamefont {V.}~\bibnamefont
      {Scarani}},\ }\href {https://doi.org/10.1093/oso/9780198788416.001.0001}
      {\emph {\bibinfo {title} {{Bell Nonlocality}}}}\ (\bibinfo  {publisher}
      {Oxford University Press},\ \bibinfo {year} {2019})\ \bibinfo {note}
      {p.38}\BibitemShut {NoStop}%
    \bibitem [{\citenamefont {Das}\ \emph {et~al.}(2022{\natexlab{a}})\citenamefont
      {Das}, \citenamefont {Karczewski}, \citenamefont {Mandarino}, \citenamefont
      {Markiewicz}, \citenamefont {Woloncewicz},\ and\ \citenamefont
      {Żukowski}}]{DasDunnComm}%
      \BibitemOpen
      \bibfield  {author} {\bibinfo {author} {\bibfnamefont {T.}~\bibnamefont
      {Das}}, \bibinfo {author} {\bibfnamefont {M.}~\bibnamefont {Karczewski}},
      \bibinfo {author} {\bibfnamefont {A.}~\bibnamefont {Mandarino}}, \bibinfo
      {author} {\bibfnamefont {M.}~\bibnamefont {Markiewicz}}, \bibinfo {author}
      {\bibfnamefont {B.}~\bibnamefont {Woloncewicz}},\ and\ \bibinfo {author}
      {\bibfnamefont {M.}~\bibnamefont {Żukowski}},\ }\bibfield  {title} {\bibinfo
      {title} {Comment on ‘{Single} particle nonlocality with completely
      independent reference states’},\ }\href
      {https://doi.org/10.1088/1367-2630/ac55b1} {\bibfield  {journal} {\bibinfo
      {journal} {New Journal of Physics}\ }\textbf {\bibinfo {volume} {24}},\
      \bibinfo {pages} {038001} (\bibinfo {year} {2022}{\natexlab{a}})}\BibitemShut
      {NoStop}%
    \bibitem [{\citenamefont {Streater}\ and\ \citenamefont
      {Wightman}(1989)}]{PCT89}%
      \BibitemOpen
      \bibfield  {author} {\bibinfo {author} {\bibfnamefont {R.~F.}\ \bibnamefont
      {Streater}}\ and\ \bibinfo {author} {\bibfnamefont {A.~S.}\ \bibnamefont
      {Wightman}},\ }\href@noop {} {\emph {\bibinfo {title} {PCT, Spin and
      Statistics, and All That}}}\ (\bibinfo  {publisher} {Princeton University
      Press},\ \bibinfo {year} {1989})\BibitemShut {NoStop}%
    \bibitem [{\citenamefont {Kitamura}(2012)}]{Kitamura12}%
      \BibitemOpen
      \bibfield  {author} {\bibinfo {author} {\bibfnamefont {T.}~\bibnamefont
      {Kitamura}},\ }\href@noop {} {\emph {\bibinfo {title} {Liquid Glass
      Transition: A Unified Theory from the Two Band Model}}}\ (\bibinfo
      {publisher} {Elsevier},\ \bibinfo {year} {2012})\ \bibinfo {note} {{Chapter}
      3}\BibitemShut {NoStop}%
    \bibitem [{\citenamefont {Eilers}\ and\ \citenamefont
      {Horst}(1975)}]{Eilers75}%
      \BibitemOpen
      \bibfield  {author} {\bibinfo {author} {\bibfnamefont {M.}~\bibnamefont
      {Eilers}}\ and\ \bibinfo {author} {\bibfnamefont {E.}~\bibnamefont {Horst}},\
      }\bibfield  {title} {\bibinfo {title} {The theorem of {Gleason} for
      nonseparable {Hilbert} spaces},\ }\href {https://doi.org/10.1007/BF01808324}
      {\bibfield  {journal} {\bibinfo  {journal} {International Journal of
      Theoretical Physics}\ }\textbf {\bibinfo {volume} {13}},\ \bibinfo {pages}
      {419} (\bibinfo {year} {1975})}\BibitemShut {NoStop}%
    \bibitem [{\citenamefont {Das}\ \emph {et~al.}(2022{\natexlab{b}})\citenamefont
      {Das}, \citenamefont {Karczewski}, \citenamefont {Mandarino}, \citenamefont
      {Markiewicz}, \citenamefont {Woloncewicz},\ and\ \citenamefont
      {Żukowski}}]{SinglePhotonGPY}%
      \BibitemOpen
      \bibfield  {author} {\bibinfo {author} {\bibfnamefont {T.}~\bibnamefont
      {Das}}, \bibinfo {author} {\bibfnamefont {M.}~\bibnamefont {Karczewski}},
      \bibinfo {author} {\bibfnamefont {A.}~\bibnamefont {Mandarino}}, \bibinfo
      {author} {\bibfnamefont {M.}~\bibnamefont {Markiewicz}}, \bibinfo {author}
      {\bibfnamefont {B.}~\bibnamefont {Woloncewicz}},\ and\ \bibinfo {author}
      {\bibfnamefont {M.}~\bibnamefont {Żukowski}},\ }\bibfield  {title} {\bibinfo
      {title} {Wave–particle complementarity: detecting violation of local
      realism with photon-number resolving weak-field homodyne measurements},\
      }\href {https://doi.org/10.1088/1367-2630/ac54c8} {\bibfield  {journal}
      {\bibinfo  {journal} {New Journal of Physics}\ }\textbf {\bibinfo {volume}
      {24}},\ \bibinfo {pages} {033017} (\bibinfo {year}
      {2022}{\natexlab{b}})}\BibitemShut {NoStop}%
    \bibitem [{\citenamefont {\ifmmode~\dot{Z}\else \.{Z}\fi{}ukowski}\ \emph
      {et~al.}(2017)\citenamefont {\ifmmode~\dot{Z}\else \.{Z}\fi{}ukowski},
      \citenamefont {Laskowski},\ and\ \citenamefont {Wie\ifmmode~\acute{s}\else
      \'{s}\fi{}niak}}]{ZUKUPRA}%
      \BibitemOpen
      \bibfield  {author} {\bibinfo {author} {\bibfnamefont {M.}~\bibnamefont
      {\ifmmode~\dot{Z}\else \.{Z}\fi{}ukowski}}, \bibinfo {author} {\bibfnamefont
      {W.}~\bibnamefont {Laskowski}},\ and\ \bibinfo {author} {\bibfnamefont
      {M.}~\bibnamefont {Wie\ifmmode~\acute{s}\else \'{s}\fi{}niak}},\ }\bibfield
      {title} {\bibinfo {title} {Normalized {Stokes} operators for polarization
      correlations of entangled optical fields},\ }\href
      {https://doi.org/10.1103/PhysRevA.95.042113} {\bibfield  {journal} {\bibinfo
      {journal} {Phys. Rev. A}\ }\textbf {\bibinfo {volume} {95}},\ \bibinfo
      {pages} {042113} (\bibinfo {year} {2017})}\BibitemShut {NoStop}%
    \bibitem [{\citenamefont {Schlichtholz}\ \emph
      {et~al.}(2022{\natexlab{a}})\citenamefont {Schlichtholz}, \citenamefont
      {Woloncewicz},\ and\ \citenamefont
      {{\.Z}ukowski}}]{schlichtholz2021simplified}%
      \BibitemOpen
      \bibfield  {author} {\bibinfo {author} {\bibfnamefont {K.}~\bibnamefont
      {Schlichtholz}}, \bibinfo {author} {\bibfnamefont {B.}~\bibnamefont
      {Woloncewicz}},\ and\ \bibinfo {author} {\bibfnamefont {M.}~\bibnamefont
      {{\.Z}ukowski}},\ }\bibfield  {title} {\bibinfo {title} {Simplified quantum
      optical {Stokes} observables and {Bell’s} theorem},\ }\href
      {https://doi.org/https://doi.org/10.1038/s41598-022-14232-8} {\bibfield
      {journal} {\bibinfo  {journal} {Scientific Reports}\ }\textbf {\bibinfo
      {volume} {12}},\ \bibinfo {pages} {1} (\bibinfo {year}
      {2022}{\natexlab{a}})}\BibitemShut {NoStop}%
    \bibitem [{\citenamefont {Schlichtholz}\ \emph
      {et~al.}(2022{\natexlab{b}})\citenamefont {Schlichtholz}, \citenamefont
      {Mandarino},\ and\ \citenamefont {Żukowski}}]{Schlichtholz_2022}%
      \BibitemOpen
      \bibfield  {author} {\bibinfo {author} {\bibfnamefont {K.}~\bibnamefont
      {Schlichtholz}}, \bibinfo {author} {\bibfnamefont {A.}~\bibnamefont
      {Mandarino}},\ and\ \bibinfo {author} {\bibfnamefont {M.}~\bibnamefont
      {Żukowski}},\ }\bibfield  {title} {\bibinfo {title} {Bosonic fields in
      states with undefined particle numbers possess detectable non-contextuality
      features, plus more},\ }\href {https://doi.org/10.1088/1367-2630/ac919e}
      {\bibfield  {journal} {\bibinfo  {journal} {New Journal of Physics}\ }\textbf
      {\bibinfo {volume} {24}},\ \bibinfo {pages} {103003} (\bibinfo {year}
      {2022}{\natexlab{b}})}\BibitemShut {NoStop}%
    \bibitem [{\citenamefont {Schlichtholz}\ \emph {et~al.}(2021)\citenamefont
      {Schlichtholz}, \citenamefont {Woloncewicz},\ and\ \citenamefont
      {\ifmmode~\dot{Z}\else \.{Z}\fi{}ukowski}}]{BGHZ}%
      \BibitemOpen
      \bibfield  {author} {\bibinfo {author} {\bibfnamefont {K.}~\bibnamefont
      {Schlichtholz}}, \bibinfo {author} {\bibfnamefont {B.}~\bibnamefont
      {Woloncewicz}},\ and\ \bibinfo {author} {\bibfnamefont {M.}~\bibnamefont
      {\ifmmode~\dot{Z}\else \.{Z}\fi{}ukowski}},\ }\bibfield  {title} {\bibinfo
      {title} {Nonclassicality of bright {Greenberger}-{Horne}-{Zeilinger}--like
      radiation of an optical parametric source},\ }\href
      {https://doi.org/10.1103/PhysRevA.103.042226} {\bibfield  {journal} {\bibinfo
       {journal} {Phys. Rev. A}\ }\textbf {\bibinfo {volume} {103}},\ \bibinfo
      {pages} {042226} (\bibinfo {year} {2021})}\BibitemShut {NoStop}%
    \bibitem [{\citenamefont {Paterek}\ \emph {et~al.}(2011)\citenamefont
      {Paterek}, \citenamefont {Kurzyński}, \citenamefont {Oi},\ and\
      \citenamefont {Kaszlikowski}}]{Paterek11}%
      \BibitemOpen
      \bibfield  {author} {\bibinfo {author} {\bibfnamefont {T.}~\bibnamefont
      {Paterek}}, \bibinfo {author} {\bibfnamefont {P.}~\bibnamefont {Kurzyński}},
      \bibinfo {author} {\bibfnamefont {D.~K.~L.}\ \bibnamefont {Oi}},\ and\
      \bibinfo {author} {\bibfnamefont {D.}~\bibnamefont {Kaszlikowski}},\
      }\bibfield  {title} {\bibinfo {title} {Reference frames for {Bell} inequality
      violation in the presence of superselection rules},\ }\href
      {https://doi.org/10.1088/1367-2630/13/4/043027} {\bibfield  {journal}
      {\bibinfo  {journal} {New Journal of Physics}\ }\textbf {\bibinfo {volume}
      {13}},\ \bibinfo {pages} {043027} (\bibinfo {year} {2011})}\BibitemShut
      {NoStop}%
    \bibitem [{\citenamefont {Das}\ \emph {et~al.}(2021)\citenamefont {Das},
      \citenamefont {Karczewski}, \citenamefont {Mandarino}, \citenamefont
      {Markiewicz}, \citenamefont {Woloncewicz},\ and\ \citenamefont
      {Żukowski}}]{firstSinglePhoton}%
      \BibitemOpen
      \bibfield  {author} {\bibinfo {author} {\bibfnamefont {T.}~\bibnamefont
      {Das}}, \bibinfo {author} {\bibfnamefont {M.}~\bibnamefont {Karczewski}},
      \bibinfo {author} {\bibfnamefont {A.}~\bibnamefont {Mandarino}}, \bibinfo
      {author} {\bibfnamefont {M.}~\bibnamefont {Markiewicz}}, \bibinfo {author}
      {\bibfnamefont {B.}~\bibnamefont {Woloncewicz}},\ and\ \bibinfo {author}
      {\bibfnamefont {M.}~\bibnamefont {Żukowski}},\ }\bibfield  {title} {\bibinfo
      {title} {Can single photon excitation of two spatially separated modes lead
      to a violation of {Bell} inequality via weak-field homodyne measurements?},\
      }\href {https://doi.org/10.1088/1367-2630/ac0ffe} {\bibfield  {journal}
      {\bibinfo  {journal} {New Journal of Physics}\ }\textbf {\bibinfo {volume}
      {23}},\ \bibinfo {pages} {073042} (\bibinfo {year} {2021})}\BibitemShut
      {NoStop}%
    \bibitem [{\citenamefont {Busch}\ \emph {et~al.}(1991)\citenamefont {Busch},
      \citenamefont {Lahti},\ and\ \citenamefont {Mittelstaedt}}]{Busch91}%
      \BibitemOpen
      \bibfield  {author} {\bibinfo {author} {\bibfnamefont {P.}~\bibnamefont
      {Busch}}, \bibinfo {author} {\bibfnamefont {P.~J.}\ \bibnamefont {Lahti}},\
      and\ \bibinfo {author} {\bibfnamefont {P.}~\bibnamefont {Mittelstaedt}},\
      }\bibinfo {title} {The quantum theory of measurement},\ in\ \href
      {https://doi.org/10.1007/978-3-662-13844-1_3} {\emph {\bibinfo {booktitle}
      {The Quantum Theory of Measurement}}}\ (\bibinfo  {publisher} {Springer
      Berlin Heidelberg},\ \bibinfo {address} {Berlin, Heidelberg},\ \bibinfo
      {year} {1991})\ pp.\ \bibinfo {pages} {27--98}\BibitemShut {NoStop}%
    \bibitem [{\citenamefont {Zurek}(1982)}]{Zurek.82}%
      \BibitemOpen
      \bibfield  {author} {\bibinfo {author} {\bibfnamefont {W.~H.}\ \bibnamefont
      {Zurek}},\ }\bibfield  {title} {\bibinfo {title} {Environment-induced
      superselection rules},\ }\href {https://doi.org/10.1103/PhysRevD.26.1862}
      {\bibfield  {journal} {\bibinfo  {journal} {Phys. Rev. D}\ }\textbf {\bibinfo
      {volume} {26}},\ \bibinfo {pages} {1862} (\bibinfo {year}
      {1982})}\BibitemShut {NoStop}%
    \bibitem [{\citenamefont {Zurek}(2003)}]{Zurek.03}%
      \BibitemOpen
      \bibfield  {author} {\bibinfo {author} {\bibfnamefont {W.~H.}\ \bibnamefont
      {Zurek}},\ }\bibfield  {title} {\bibinfo {title} {Decoherence, einselection,
      and the quantum origins of the classical},\ }\href
      {https://doi.org/10.1103/RevModPhys.75.715} {\bibfield  {journal} {\bibinfo
      {journal} {Rev. Mod. Phys.}\ }\textbf {\bibinfo {volume} {75}},\ \bibinfo
      {pages} {715} (\bibinfo {year} {2003})}\BibitemShut {NoStop}%
    \end{thebibliography}
\end{document}